\documentclass[12pt,a4paper]{article}
\pdfoutput=1 %Required by JCAP,JHEP
\usepackage{amsmath,amssymb,graphics,graphicx,epstopdf,color,hyperref,epsfig}
\def\be{\begin{equation}}
\def\ee{\end{equation}}
\def\bea{\begin{eqnarray}}
\def\eea{\end{eqnarray}}
\usepackage{graphicx}% Include figure files

\newcommand{\nn}{\nonumber}

\def\beq{\begin{eqnarray}}
\def\eeq{\end{eqnarray}}
\def\nn{\nonumber}
\def\lam{\lambda}

\catcode`\@=11
\def\lsim{\mathrel{\mathpalette\@versim<}}
\def\gsim{\mathrel{\mathpalette\@versim>}}
\def\@versim#1#2{\vcenter{\offinterlineskip
\ialign{$\m@th#1\hfil##\hfil$\crcr#2\crcr\sim\crcr } }}
\catcode`\@=12

\catcode`@=12
\begin{document}
\thispagestyle{empty}
\begin{flushright}
UCRHEP-T559\\
December 2015\
\end{flushright}
\vspace{0.6in}
\begin{center}
{\LARGE \bf Radiative Model of Neutrino Mass with\\ 
Neutrino Interacting MeV Dark Matter\\}
\vspace{1.0in}
{Abdesslam Arhrib$^1$, C${\rm \acute{e}}$line B{\oe}hm$^2$, Ernest Ma$^3$ and 
Tzu-Chiang Yuan$^{4,5}$\\}
\vspace{0.5in}
{\it $^1$D${\it \acute{e}}$partement de Math${\it \acute{e}}$matique, 
Facult${\it \acute{e}}$ des Sciences et Techniques,\\}
{\it Universit${\it \acute{e}}$ Abdelmalek Essaadi, B. 416, Tangier, Morocco\\}
{\it $^2$LAPTH, UMR 5108, 9 chemin de Bellevue - BP 110, 74941 
Annecy-Le-Vieux, France\\}
{\it $^3$Department of Physics and Astronomy,\\}
%\vspace{0.1in}
{\it University of California, 
Riverside, California 92521, USA\\}
{\it $^4$Institute of Physics, Academia Sinica, Nangang, Taipei 11529, Taiwan\\}
{\it $^5$Physics Division, National Center for Theoretical Sciences, Hsinchu, Taiwan}
\end{center}
\vspace{0.5in}

\begin{abstract}
We consider the radiative generation of neutrino mass through the interactions 
of neutrinos with MeV dark matter.  We construct a realistic renormalizable 
model with one scalar doublet (in additional to the standard model doublet) 
and one complex singlet together with three 
light singlet Majorana fermions, all transforming under a dark $U(1)_D$ 
symmetry which breaks softly to $Z_2$.  We study in detail the scalar sector 
which supports this specific scenario and its rich phenomenology. 
\end{abstract}

\newpage
\baselineskip 24pt

\section{Introduction}

The nature of dark matter (DM) is one of the most disputed topics in 
cosmology.  Until one (or two) decade(s) ago, only a few candidates  
prevailed in the literature, among which were  neutralinos (a thermal, cold, DM 
species) and axions (also cold DM but non-thermal).  Astrophysical and cosmological 
anomalies since in the last 10-15 years however led many authors to study more exotics scenarios, such as light DM, leptophilic DM, 
sterile neutrinos \cite{Boehm:2003bt,Cirelli:2005uq,Fox:2008kb,Goodenough:2009gk,Jeltema:2014qfa}.  
So far most DM studies have focused on either almost massless particles (axions), keV particles (sterile neutrinos) or GeV to TeV DM candidates (as provided by supersymmetry and Kaluza-Klein theories\footnote{For a review see Ref.~\cite{Bertone:2004pz}.}) but the range between a few keV and GeV has been somewhat disregarded. 

In cosmology, both keV and GeV-TeV DM candidates are generally assumed to be collisionless, even though their annihilations or decay are invoked to explain the observed DM abundance.  
About a decade ago, it was pointed out that -- even weak-strength -- DM 
interactions could erase the DM primordial interactions and should not  be disregarded when the DM is relatively light (a few MeV) and coupled to neutrinos or photons \cite{Boehm:2000gq,Boehm:2004th}. Indeed the damping of the DM primordial fluctuations has two origins, as shown in these references: one is the collisional damping, which 
suppresses the matter fluctuations until the DM is kinetically decoupled from any 
other species; this is analogous to the Silk damping. The other source is the DM free-streaming which erases fluctuations that have not been erased yet by the DM collisions.

The resulting linear matter power spectrum associated with  light DM candidates coupled to radiation features  damped oscillations in addition to an exponential cut-off \cite{Boehm:2001hm,Sigurdson:2003vy,Mangano:2006mp}. This makes these scenarios  interesting alternatives to vanilla CDM and Warm DM candidates.

There has been much interest in the DM-neutrino coupling since these first studies but with the twist of DM self-interactions \cite{Hannestad:2013ana,Chu:2015ipa, Mirizzi:2014ama, Bilenky:1999dn}.   However, as shown in Refs.~\cite{Boehm:2014vja,Schewtschenko:2014fca,Schewtschenko:2015rno}, a sole DM-neutrino coupling can solve the missing satellite (which is a deficit of dwarf galaxy haloes in Milky Way-like DM haloes) and the  too big to fail problems when the DM elastic scattering off neutrinos is of the order of 
\begin{equation}
 \sigma_{\rm el} \simeq 10^{-36} \ \left( \frac{m_{\rm{DM}}}{\rm{MeV}}\right) \  \rm{cm^2}.
 \label{sigmael_i} 
\end{equation}
For  DM candidates with a mass of about a few MeV, these interactions are typically of the order of the  Standard Model weak interactions. Assuming a simple crossing between the elastic scattering  and annihilations processes, one expects 
an annihilation cross section of the order of
\begin{equation}
\sigma v \simeq 3\cdot  10^{-26} \ \left( \frac{m_{\rm{DM}}}{\rm{MeV}}\right) \  \rm{cm^3/s},
 \label{sigmav_i} 
\end{equation}
which is the  required value to explain the observed DM abundance in thermal DM scenarios. 
The correspondence between Eq.~(\ref{sigmael_i})  and Eq.~(\ref{sigmav_i}) thus suggests that current cosmological problems could be related to the current DM abundance.

Even more puzzling is the possibility to explain in some specific models \cite{Boehm:2006mi,Farzan:2009,Farzan:2010mr} the existence of small neutrino masses in the presence of such a DM-neutrino coupling. It is therefore tempting to assume that there exists a framework in which DM-neutrino interactions can explain simultaneously the missing satellite and too big to fail problems, the existence of small neutrinos masses and the observed DM abundance.

In this paper we construct such a framework. We envision a fundamental Yukawa coupling 
of the form $\bar{N}_R \nu_L \zeta_2$ where the dark matter candidate, here referred to as $N$, 
is a Majorana fermion and  both the fermion $N$ and the 
scalar $\zeta_2$ are light, with masses of order a few MeV.  
In Section 2, we review the elastic scattering cross section 
among the neutrino and DM and the related process of DM annihilation 
into neutrino pair based on this Yukawa coupling.
To support this specific scenario, we study an extension of the standard 
model with one additional scalar doublet and one additional complex singlet, 
both of which transform nontrivially under a dark global $U(1)_D$ that is softly broken into a discrete $Z_2$ (Section 3). 
We show how realistic neutrino masses may be obtained with a scalar mass 
spectrum including the light $\zeta_2$ without conflicting with present  
data at the Large Hadron Collider (LHC) (Sections 3 and 4).  We examine also in detail the scalar 
sector and obtain theoretical and phenomenological constraints on its 
parameter space (Section 4). Numerical results are presented in Section 5 and conclusion in Section 6.
Some useful formulas are collected in the Appendix.

\section{Elastic scattering and annihilation cross sections}

In (thermal) scenarios where  DM can scatter off neutrinos, the collisional damping scale is determined by the integral 
\begin{equation}
l^2_{\rm{coll \ damping}}
\simeq \int^{t_{{\rm dec}({\rm DM}-\nu)}} \frac{\rho_{\nu} }{(\rho+p)_{\rm tot} \ a^2 \ \Gamma_\nu} \ v^2 \ d t \; , 
\end{equation}
where $a$ is the scale factor, $\rho_{\nu}$ the neutrino energy density, $\Gamma_{\nu}$ the neutrino interaction rate, $v$ the neutrino velocity, $(\rho+p)_{\rm tot} $ is the sum over the energy density and pressure of all the species coupled to the DM while DM still interacts with neutrinos (which includes the DM itself).   
This length is  directly proportional to the neutrino density and velocity (which is equal to $c$ if one assumes that the DM kinetic decoupling from neutrinos happens well before neutrinos become non relativistic) and the neutrino kinetic decoupling time \cite{Boehm:2000gq,Boehm:2004th}.  Its magnitude also depends on the period over which the DM is coupled to neutrinos; hence the integral over time, with $t_{\rm{dec}({\rm DM}-\nu)}$ (the DM decoupling time from neutrino) as upper bound. 

The CMB and linear matter power spectra in the presence of such a DM-neutrino coupling can be easily predicted using the Boltzmann formalism~\cite{Mangano:2006mp, 
Wilkinson:2014ksa}. Both agree with the damping length estimate obtained analytically using the above formula (in the absence of mixed damping). But the matter power spectrum ultimately sets the stronger constraint, namely 
\begin{equation}
\sigma_{\rm el} = 10^{-36} \ \left(\frac{m_{\rm{DM}}}{\rm{MeV}} \right)  \ \rm{cm^2} \; ,
\end{equation}
if the cross section is independent of the temperature or
\begin{equation}
\sigma_{\rm el} = 10^{-48} \ \left(\frac{m_{\rm{DM}}}{\rm{MeV}} \right)  \ \left(\frac{T}{\rm{2.7 \cdot 10^{-4} \ eV}}\right)^2   \ \rm{cm^2} \; , 
\end{equation}
if the cross section depends on the neutrino energy   \cite{Wilkinson:2014ksa}. This confirms that a weak strength cross section can erase DM fluctuations at cosmologically relevant scales, if the DM is relatively light. The simplest elastic scattering process $N\ \nu \rightarrow N \ \nu $ that gives rise to such an effect relies on the exchange of a fermion (scalar) if the DM is a scalar (fermion).  The cross section  for a Majorana candidate coupled to neutrinos with a coupling $g$ 
is  given by the u and s-channels diagrams, leading to:
\begin{equation}
\sigma_{\rm el} \simeq \frac{3 \ g^4 }{16 \ \pi} \ \frac{T^2}{(m_{N}^2-m_{\zeta_2}^2)^2} \; , 
\label{sigma_el}
\end{equation}
in the absence of a close  degeneracy between the mediator and DM masses. Here we also implicitly assume MeV DM, 
{\it i.e.} that DM is non relativistic at the DM-neutrino decoupling, which occurs slightly below a keV. The annihilation diagrams (t and u-channels) lead to the dominant contribution
\begin{equation} 
\sigma v \simeq  \frac{g^4}{4 \ \pi  \   m_{\rm{\zeta_2}}^2} \simeq 2.38 \cdot 10^{-26} \  \left(\frac{g}{4 \cdot 10^{-4}}\right)^4 \   \left(\frac{m_{\rm{\zeta_2}}}{\rm{MeV}}\right)^{-2}  \rm{cm^3/s} \; ,
\label{sigma_ann}
\end{equation}
where we again assume that there is not a strict degeneracy between the DM and mediator masses and neglect the neutrino mass.

Eqs.~(\ref{sigma_el}) and (\ref{sigma_ann}) cannot be satisfied simultaneously with the same values of the mass and couplings, unless the DM mass is slightly smaller than a few keV. Yet thermal keV annihilating DM particles into neutrinos are already ruled out \cite{Serpico:2004nm, Boehm:2012gr, Boehm:2013jpa, Nollett:2014lwa}, as they would change the number of relativistic  degrees of  freedom at nucleosynthesis and  CMB time, by too large an amount. The only possibility for thermal DM candidates coupled to neutrinos is to have a mass above a few MeV \cite{Boehm:2013jpa}.

In order to explain both the DM abundance and solve cosmological problems, one 
thus needs thus to get rid off the temperature dependence of the elastic scattering cross section. 
This occurs if the mass splitting between $N$ and $\zeta_2$ are of the order of a few keV or below.  
Indeed in this case the elastic scattering cross section reads
\begin{equation}
 \sigma_{\rm el} \simeq  \frac{g^4}{16 \ \pi \ m_{N}^2} \simeq 10^{-36} \ \left(\frac{g}{6 \cdot 10^{-4}}\right)^4 \   \left(\frac{m_{\rm{N}}}{\rm{MeV}}\right)^{-2}  \ \rm{cm^2}  \; ,
 \label{sigma_elworks}
\end{equation}
while the annihilation  cross section is given by
\begin{equation}
\sigma v \simeq  \frac{g^4 \ m_N^2}{4 \ \pi  \   (m_{\rm{\zeta_2}}^2 + m_{N}^2)^2} \simeq 3 \cdot 10^{-26} \  
\left(\frac{g}{6 \cdot 10^{-4}}\right)^4 \   \left(\frac{m_{\rm{N}}}{\rm{MeV}}\right)^{-2}  \rm{cm^3/s} \; .
 \label{sigma_annworks}
\end{equation}
Therefore, a scenario where the DM is of a few MeVs but the mediator is only slightly heavier than the DM by a few keVs  can  solve the missing satellite and too big to fail problems (in the absence of baryonic physics) and also explain the DM observed abundance.

Note that the presence of baryonic interactions could alter these values. Depending on the magnitude of the effect, one might either lose the above correspondence or be able to make a temperature dependent elastic scattering and temperature independent annihilation cross section compatible. Given that such studies do not exist yet, we will take the above numbers (see Eqs.~(\ref{sigma_elworks}) and (\ref{sigma_annworks})) at face value.   

We now investigate whether such a scenario can also give rise to neutrino masses. 
Note that such a scenario predicts a slightly larger value of $N_{\rm eff}$ than 3.046 and $H_0 \simeq \ \rm{71 \ km/s/Mpc}$ \cite{Wilkinson:2014ksa}.

\section{Radiative Neutrino Mass Through Dark Matter}

The simplest finite one-loop radiative model of neutrino mass through dark 
matter is the scotogenic model (from the Greek ``scotos'' meaning darkness) 
proposed in 2006~\cite{m06}.  It assumes an exactly conserved $Z_2$ 
symmetry~\cite{dm78} and extends the standard model (SM) of particle 
interactions with the addition of one 
scalar doublet $(\eta^+,\eta^0)$ and three singlet Majorana fermions 
$N_{1,2,3}$ which are odd under $Z_2$.  Many aspects of its phenomenology 
have been studied in detail~\cite{atyy14}.  Whereas the masses of $\eta$ 
and $N$ are usually considered to be heavy, this mechanism also allows $N$ to 
be light~\cite{m12}.  With the discovery~\cite{atlas12,cms12} of the 125 
GeV particle at the Large Hadron Collider (LHC) and its identification 
with the long-sought Higgs boson $h$ of the SM, important constraints 
on $\eta$ are now applicable.  From the limit on the invisible width of $h$, 
a light scalar ($\sim$ MeV) is not allowed in the context of the original 
scotogenic model. 

In this paper we consider the further addition of a complex scalar singlet 
$\chi$ and impose a dark $U(1)_D$ symmetry which is softly broken to $Z_2$. 

Whereas all SM fields have zero $U(1)_D$ charge, all 
other fields $\eta$, $\chi$ and $N_{1,2,3}$ are assumed to have the same 
nonzero $U(1)_D$ charge, say +1.  To get the required neutrino mass and 
Higgs interaction, it does not work with just the inert Higgs doublet $\eta$, 
nor with the addition of a real scalar singlet.  However, as shown in this 
paper, it will work with the $\eta$ doublet plus a complex singlet $\chi$, 
which is naturally maintained with a dark $U(1)_D$ symmetry, softly broken 
to $Z_2$ so that $N$ may have a Majorana mass and $\nu$ gets a one-loop 
radiative mass as shown in Fig.~1.  If only $Z_2$ is used, then many more 
allowed terms appear in the Lagrangian, such as $(\Phi^\dagger \Phi) \chi^2 + 
{\rm H.c.}$ on top of $(\Phi^\dagger \Phi) \chi^* \chi$, which are unnecessary 
complications (and must indeed be small) in the subsequent discussion of 
the scalar sector.

The scalar potential is given by
\begin{eqnarray}
V &=& m_1^2 \Phi^\dagger \Phi + m_2^2 \eta^\dagger \eta + m_3^2 
\chi^* \chi + \frac{1}{2} m_4^2 [\chi^2 + (\chi^*)^2] \nonumber \\ 
&+& \mu [\eta^\dagger \Phi \chi + \Phi^\dagger \eta \chi^*] 
+ \frac{1}{2} \lambda_1 (\Phi^\dagger \Phi)^2 
+ \frac{1}{2} \lambda_2 (\eta^\dagger \eta)^2 
+ \frac{1}{2} \lambda_3 (\chi^* \chi)^2 \nonumber \\
&+& \lambda_4 (\eta^ \dagger \eta)(\Phi^\dagger \Phi) 
+ \lambda_5 (\eta^ \dagger \Phi)(\Phi^\dagger \eta) 
+ \lambda_6 (\chi^* \chi)(\Phi^\dagger \Phi)
+ \lambda_7 (\chi^* \chi)(\eta^\dagger \eta) \; ,
\label{eq:pot}
\end{eqnarray}
where the $m_4^2$ term breaks $U(1)_D$ softly to $Z_2$.  Note that the quartic 
term $(\Phi^\dagger \eta)^2$ present in the original $Z_2$ model~\cite{m06} is 
now forbidden.  

\begin{figure}[hptb]
\includegraphics[width=0.990\textwidth]{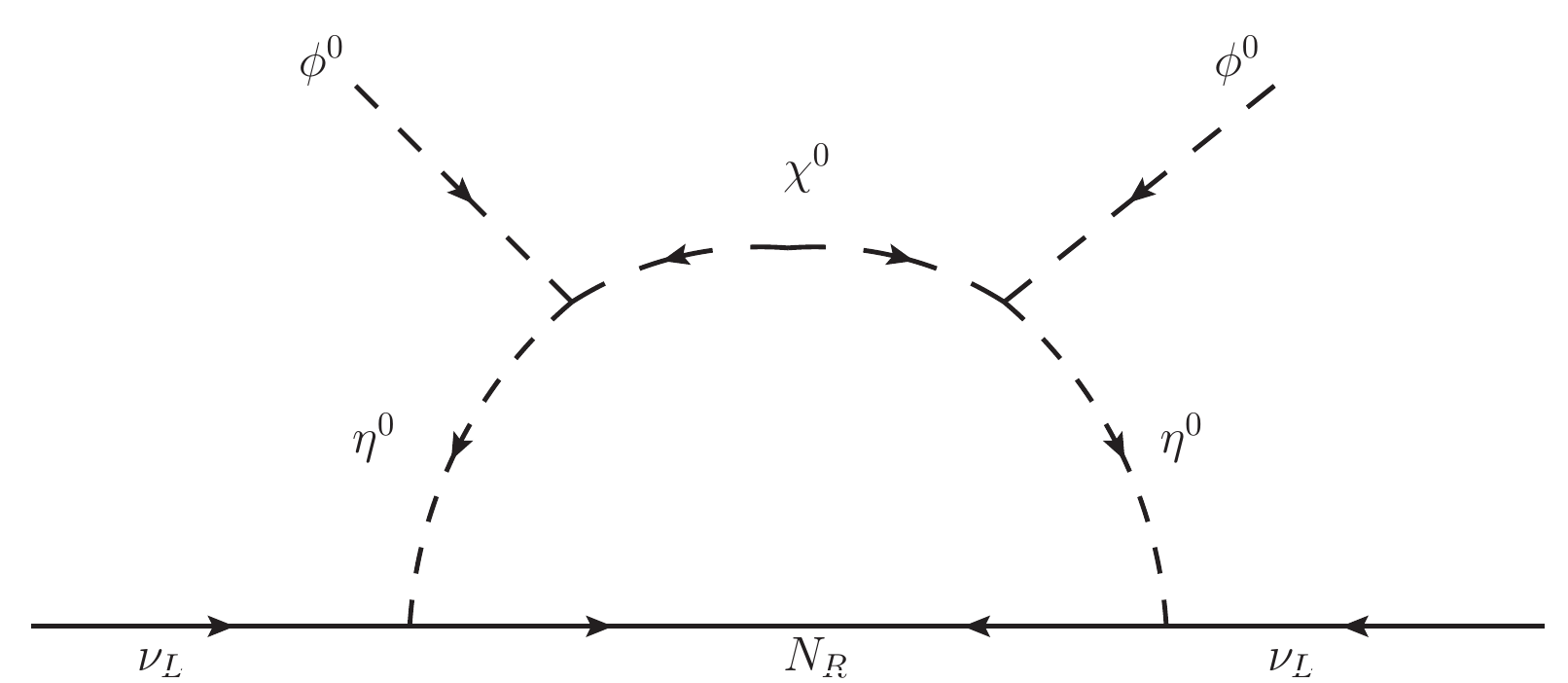}
\caption{One-loop scotogenic neutrino mass from $U(1)_D$ breaking to $Z_2$.}
\end{figure}

The one-loop mechanism for neutrino mass is depicted in 
Fig.~1, from which we can easily see that 
the Majorana mass term for $N_i$   
breaks $U(1)_D$ to $Z_2$.
This diagram is similar to that required in a supersymmetric 
extension~\cite{m06-1}.
We note that the $m_4^2$ and Majorana mass terms 
are the only two terms in the model that softly break $U(1)_D$ into $Z_2$.  
This is a well-known method in symmetry breaking,
because soft breaking generates only finite corrections in the
renormalizable terms of the Lagrangian and a well-known rationale for
setting $m_4$ small against other mass parameters (because its absence
enlarges the symmetry of the theory).  
The model however remains renormalizable.  
These soft terms are analogous to the sfermion mass and the gaugino mass in MSSM
which break supersymmetry softly. The origin of these softly breaking terms
may be revealed only at a higher theory. In the case of MSSM, these soft terms
could arise from supergravity. In our case, we will treat them as phenomenological terms,
just like the superpartner mass terms in MSSM, without worrying about the higher theory.
For an early application of this idea in neutrino physics, see for example~\cite{Ma:2000cc}.

Let $\Phi = [\omega_1^+, (v+\phi_R+i\phi_I)/\sqrt{2}]^T$, 
$\eta = [\eta^+, (\eta_R + i \eta_I)/\sqrt{2}]^T$ and
$\chi = (\chi_R + i \chi_I)/\sqrt{2}$.
The mass of the SM-like Higgs $h$ ($\equiv \phi_R$) and charged Higgs $\eta^\pm$ are given by:
\begin{eqnarray}
m_h^2 &=&  \lambda_1 v^2 \; , \\
m_{\eta^\pm}^2 &= & m_2^2 + \frac{1}{2}\lambda_4 v^2 \; .
\label{eq:mass1}
\end{eqnarray}

The neutral components of $\eta$ and $\chi$ will mix 
through $\mu$ term of the potential. The mass-squared matrices 
spanning $(\eta_{R,I},\chi_{R,I})$ are given by
\begin{eqnarray}
{\cal M}^2_{R,I}=
\left(
\begin{array}{cc}
m_2^2 + (\lambda_4 + \lambda_5) v^2/2  & \mu v/\sqrt{2}  \\
\mu v/\sqrt{2}  & m_3^2 + \lambda_6 v^2/2 \pm m_4^2
\end{array}\right) \; .
\label{eq:mix}
\end{eqnarray}

Let $\zeta_{1R}, \zeta_{2R}, \zeta_{1I}, \zeta_{2I}$ be the mass eigenstates 
with masses $m_{1R}, m_{2R}, m_{1I}, m_{2I}$:
\begin{eqnarray}
&& \zeta_{1R} = \cos \theta_{R} \eta_R - \sin \theta_R \chi_R, ~~~ 
\zeta_{2R} = \sin \theta_{R} \eta_R + \cos \theta_R \chi_R, \\
&& \zeta_{1I} = \cos \theta_{I} \eta_I - \sin \theta_I \chi_I, ~~~ 
\zeta_{2I} = \sin \theta_{I} \eta_I + \cos \theta_I \chi_I, 
\end{eqnarray}
then the neutrino mass matrix is given by
\begin{eqnarray}
({\cal M}_\nu)_{ij} &=& \sum_k \frac{h_{ik} h_{jk} M_k}{16 \pi^2} 
\left[ \frac{\cos^2 \theta_R m^2_{1R}}{m^2_{1R} - M_k^2} \ln \frac{m^2_{1R}}{M_k^2} 
+ \frac{\sin^2 \theta_R m^2_{2R}}{m^2_{2R} - M_k^2} \ln \frac{m^2_{2R}}{M_k^2} 
\right. \nonumber \\ 
&& \left. - \frac{\cos^2 \theta_I m^2_{1I}}{m^2_{1I} - M_k^2} 
\ln \frac{m^2_{1I}}{M_k^2} - \frac{\sin^2 \theta_I m^2_{2I}}{m^2_{2I} - M_k^2} 
\ln \frac{m^2_{2I}}{M_k^2} \right],
\end{eqnarray}
where $M_k$ are the masses of $N_k$.  Note that in the limit of $m_4^2 = 0$, 
$m_{1R} = m_{1I}$, $m_{2R} = m_{2I}$, and $\theta_R = \theta_I$.  Hence the 
neutrino mass would be zero.

We assume that $m_4^2$ is very small, then
\begin{eqnarray}
&& m^2_{1R} = m_{10}^2 + s^2 m_4^2, ~~~ m^2_{1I} = m_{10}^2 - s^2 m_4^2, \\ 
&& m^2_{2R} = m_{20}^2 + c^2 m_4^2, ~~~ m^2_{2I} = m_{20}^2 - c^2 m_4^2, \\ 
&& \sin \theta_R = s \left( 1 + \frac{c^2 m_4^2}{m_{10}^2 - m_{20}^2} \right), 
~~~ \sin \theta_I = s \left( 1 - \frac{c^2 m_4^2}{m_{10}^2 - m_{20}^2} \right), 
\\ 
&& \cos \theta_R = c \left( 1 -  \frac{s^2 m_4^2}{m_{10}^2 - m_{20}^2} \right), 
~~~ \cos \theta_I = c \left( 1 + \frac{s^2 m_4^2}{m_{10}^2 - m_{20}^2} \right), 
\end{eqnarray}
where $s = \sin \theta_0$ and $c = \cos \theta_0$ which diagonalize 
the $(\eta^0,\chi)$ mass-squared matrix in the absence of $m_4^2$ with 
eigenvalues $m^2_{10}$ and $m^2_{20}$.  
The one-loop neutrino mass matrix is then of the form
\begin{equation}
({\cal M}_\nu)_{ij} = \frac{ s^2 c^2 m_4^2}{8 \pi^2} \sum_k h_{ik} h_{jk} 
M_k \left[ \frac{1 - 2 \ln(m_{10}^2/M_k^2)}{m_{10}^2 - M_k^2} -  
\frac{1 - 2 \ln(m_{20}^2/M_k^2)}{m_{20}^2 - M_k^2} \right] \; .
\end{equation}
For $m_{10}$ of order 100 GeV and $m_{20},M$ of order MeV, the first term 
is negligible.  For example, let $h{\rm s} = 0.2$, $s = 0.5$, $M = 3$ MeV, 
$m_{20} = 4$ MeV, and $m_4^2 = (128{~\rm keV})^2$, then $m_\nu = 0.1$ eV.
In this scenario, $N$ is dark matter with a mass of 3 MeV, $\zeta_2$ 
has a mass of 4 MeV and interacts with $\bar{\nu}_L N_R$ with strength 0.1. 
This is thus a possible scenario for neutrino interacting MeV dark 
matter which obtains the correct relic abundance, as discussed in Section 2.  
Note that the $\zeta_2$ mass splitting 
is small, {\it i.e.} 3 keV, and both $\zeta_{2R}$ and $\zeta_{2I}$ decay to 
$\nu N$.

If the cosmological missing-satellite problem and the too-big-to-fail 
problem are solved using elastic $N \nu$ scattering, then $\zeta_{2R}$ 
and $\zeta_{2I}$ should be both only a few keV above $M$, in which case  
Eq.~(21) is not valid.  Let $m_{2R}^2 = M^2 (1 + \delta_R)$ and 
$m_{2I}^2 = M^2 (1 + \delta_I)$, with $\delta_{R,I}$ of order $10^{-3}$, 
then the radiative neutrino mass becomes $(s^2 h^2/32 \pi^2) M (\delta_I 
- \delta_R)$, which is of order 0.1 eV for $h$s = 0.1 as desired.

However, for the small couplings implied by Eqs.~(\ref{sigma_elworks}) and (\ref{sigma_annworks}), 
the induced neutrino mass is negligible.  On the other hand, only $N_1$ needs to
be light, whereas $N_{2,3}$ can be heavy and the usual acceptable
neutrino masses are obtained.  The important point of this study (to be justified in the subsequent sections) is that the mass of one scalar,
{\it i.e.} $\zeta_2$ ($\zeta_{2R}$ or $\zeta_{2I}$), can be of order MeV. The other two neutral scalars $\zeta_{1R}$ and $\zeta_{1I}$ 
can be heavy. We will assume $\zeta_{1R}$ and $\zeta_{1I}$ are heavier than $m_h /2$ so that they do not contribute to the invisible Higgs width.

%%%%%%%%%%%%%%%%%%%%%%%%%%%%%%%%%%%
\section{Phenomenology of the Scalar Sector: 
         Theoretical and Experimental Constraints}

The scalar potential Eq.~(\ref{eq:pot}) has altogether 12 parameters and 1 vacuum expectation value (vev) $v$. 
Two of them ($m_1^2$ and $v$) can be eliminated
by  the minimization condition and $W$ gauge boson mass. 
At the LHC, both ATLAS and CMS experiments had performed several measurements 
of the newly discovered scalar particle in different channels.
The combined measured mass performed by ATLAS and CMS collaborations based on the data from 
$h\to \gamma \gamma$ and $h\to ZZ\to 4l$ channels is 
$m_{h} =$ 125.09 $\pm$ 0.21 (stat.) $\pm$ 0.11 (syst.) GeV~\cite{Aad:2015zhl}.
This measurement if interpreted as the SM Higgs boson 
allow us to fix $\lambda_1$. 
We are then left with 10 independent parameters:
\begin{eqnarray}
{\cal P} = \{ \lambda_{2,3,4,5,6,7} , m_{2}^2, m_3^2 , m_4^2, \mu  \} \; .
\end{eqnarray}
In our numerical analysis presented in the next Section, 
these parameters are scanned in the confined domain that fulfill various
theoretical and experimental constraints which are discussed below.

\subsection{Theoretical Constraints}

%%%%%%%%%%%%%%%%%%%%%%%%%%%%%%%%%%%
\subsubsection{Unitarity Constraints}

Our scalar potential is similar to the one in the 2 Higgs doublet model 
except augmented by a complex singlet field $\chi$.
We can carefully use the full set of unitarity constraints derived for the 
2 Higgs doublet model in \cite{Akeroyd:2000wc}.
In Appendix A.1, we list the set of unitarity constraints 
that we will use. Some of the $2 \to 2$ scattering amplitudes have been modified to
take into account the presence of $\chi$.
In summary, the requirement that the largest eigenvalues of all the partial wave matrices $a_0s$
for different channels to respect the unitarity constraints implies
\begin{eqnarray}
|a_{\pm}| \ , \ |b_{\pm}| \ , \ |c_{\pm}|  \ , \  |s_{1,2}| \ , \ 
|f_{\pm}| \ , \ |e_{1,2}| \ , \ |f_{1,2}| \ , \ |p_{1}| \ \leq 8\pi \; ,
\label{constraint}
\end{eqnarray}
where the definitions of the eigenvalues ($a_{\pm}$, $b_{\pm}$, and so on in the above equation) 
in terms of the quartic couplings in the scalar potential
can be found in Appendix A.1.

%%%%%%%%%%%%%%%%%%%%%%%%%%%%%%%%%%%
\subsubsection{Vacuum Stability}

A necessary condition for the stability of the vacuum comes 
from requiring the potential given
in Eq.~(\ref{eq:pot}) to be bounded from below when the scalar 
fields become large in any direction of the field space.  
At large field values, the scalar potential is dominated by quartic
couplings, the bounded from below constraints will depend only on the
quartic couplings.
The constraints ensuring tree level vacuum stability are: 
\begin{itemize}
\item If $\lam_6>0$ and  $\lam_7>0$,
\begin{eqnarray}
\label{eq:lam67positif}
&& \lambda_1 >0 \qquad , \qquad \lambda_2 >0 \qquad , \qquad \lambda_3 >0  \; , \\
&& \sqrt{\lambda_1\lambda_2}+\lambda_4+\lambda_5>0  \; ,  \\
&& \sqrt{\lambda_1\lambda_2}+\lambda_4>0  \; .
\end{eqnarray}
\item If $\lam_6<0$ or $\lam_7<0$, in additional to the above constraints, we also have
\begin{eqnarray}
\label{eq:vac2}
&&   (\lambda_3\lambda_1-\lambda_6^2) >0 \; ,
\\
&& (\lambda_3 \lambda_2-\lambda_7^2) > 0 \; , 
\\
&&  -\lambda_6 \lambda_7+\lambda_3\lambda_4+
 \sqrt{(\lambda_3\lambda_1-\lambda_6^2) (\lambda_3 \lambda_2-\lambda_7^2)}>0  \; , 
\\
&&-\lambda_6 \lambda_7+\lambda_3(\lambda_4+\lambda_5 )+
     \sqrt{(\lambda_3\lambda_1-\lambda_6^2) (\lambda_3
       \lambda_2-\lambda_7^2)}>0  \; .
\end{eqnarray}
\end{itemize}
Additional constraints also related to the stability issue 
come from the requirement of the absence of tachyonic masses. They are
\begin{eqnarray}
\label{eq:tach1}
&& m_{\eta^+}^2=m_2^2 + \frac{1}{2} \lambda_4  v^2>0 \; ,\\
\label{eq:tach2}
&& m^2_{1R}+m^2_{2R} =m_2^2+m_3^2+m_4^2 + \frac{1}{2}(\lambda_4 + \lambda_5 +
\lambda_6) v^2>0 \; , \\
\label{eq:tach3}
&& m^2_{1I}+m^2_{2I} =m_2^2+m_3^2-m_4^2 +\frac{1}{2}(\lambda_4 + \lambda_5 +
\lambda_6) v^2>0 \; .
\end{eqnarray}
Details of derivation of these constraints can be found in Appendix A.2.

%%%%%%%%%%%%%%%%%%%%%%%%%%%%%%%%%%%

\subsection{Experimental Constraints}

\subsubsection{Invisible Decay of the Higgs}

Our neutrino model requires MeV warm dark matter particle which can be identified as the 
lightest Majorana neutrino state of $N_{1,2,3}$. 
The SM Higgs $h \to N_i N_i$ can occur only through one loop. Hence its branching ratio is small
and we will ignore this invisible mode in our analysis.
On the other hand, due to the mixing of 
complex field $\chi$ with the inert doublet $\eta$, 
we have 2 light dark Higgses $\zeta_{2R}$ and $\zeta_{2I}$, one is CP even
and one is CP odd. 
These states are not stable since they can decay via $\chi_D \to N \nu$ where 
$\chi_D$ is the lighter state of  $\zeta_{2R}$ and $\zeta_{2I}$ and $N$ is the DM, the lightest of $N_{1,2,3}$.
Thus the tree level decay $h\to\chi_D \chi_D \to NN \nu\nu$ will be invisible.
The SM Higgs couplings to these dark Higgses  $\zeta_{2R}$ and 
$\zeta_{2I}$ are given in Table 1 in Appendix A.3. 

The openings of one of the non-standard decays
of the Higgs boson such as $h\to \zeta_i \zeta_j$
can modify the total width of the Higgs boson and can have 
significant impact on LHC results.  
Both ATLAS and CMS had performed searches for invisible decay of the 
Higgs boson  \cite{higgs-invisible-atlas,higgs-invisible-cms}. 
Both experiments set upper limit 
on the branching fraction of the invisible decays of the Higgs.
These limits are of the order of 28\% from ATLAS~\cite{higgs-invisible-atlas} or 
36\% from CMS~\cite{higgs-invisible-cms}
and will be improved further with the new LHC run at 13 and 14 TeV.
This constraint on the invisible decay is still
rather weak compared to the one derived from various works of global fits to 
ATLAS and CMS data \cite{globalCLT}. 
These global fits studies with the assumption that 
the Higgs boson has SM-like couplings to all SM particles 
plus additional invisible decay mode, suggest  that the 
branching ratio of the invisible decay of the Higgs boson should not exceed  
19\% at 95\% C.L. 
In our numerical analysis presented later, we will use this global fitting result
for the invisible width of the Higgs instead of the experimental upper limits.
In our model, the SM Higgs couples to all 
SM particles like fermions, massive gauge bosons and gluons
exactly the same way as in SM. The only exceptions are $h\to \gamma\gamma$ and 
$h\to \gamma Z$ which receive additional contributions from charged Higgs.
The Higgs total width can be modified slightly by $h\to \gamma\gamma$ and 
$h\to \gamma Z$ as well as by $h\to \zeta_{iR}\zeta_{jR}$ and 
$h\to \zeta_{iI}\zeta_{jI}$ ($i,j=1,2$)  if these latter channels are open.

The couplings of the SM Higgs to the neutral dark Higgses $\eta_{R,I}$, 
$\chi_{R,I}$ is given by
\begin{eqnarray}
{\cal F}=
\left(
\begin{array}{cc}
(\lambda_4 + \lambda_5) v & \mu /\sqrt{2}  \\
\mu /\sqrt{2}  &\lambda_6 v 
\end{array}\right) \; .
\label{eq:hinv}
\end{eqnarray}
As one can see from Eq.~(\ref{eq:mix}) and 
Eq.~(\ref{eq:hinv}), if the matrices ${\cal F}$ and $ {\cal M}^2_{R,I}$  
are proportional to each other, the Higgs 
couplings to $\zeta_{2R}$ and $\zeta_{2I}$ are automatically diagonal
in the mass eigenstate basis and proportional to its mass squared.
The conditions for ${\cal F}$ and $ {\cal M}^2_{R,I}$
to be proportional to each other, in the limit of $m_4^2=0$, are
\beq
m_2^2=(\lambda_4 + \lambda_5) v^2/2\quad {\rm and}  \quad m_3^2=\lambda_6 v^2/2 \; .
\label{eq:tunem23}
\eeq
If these conditions are fulfilled, we have ${\cal M}^2_{R,I}= v {\cal F}$.
Once ${\cal M}^2_{R,I}$ are diagonalized by some orthogonal matrices, 
the coupling matrix ${\cal F}$ will be also diagonal in the mass 
eigenstate basis. Therefore 
the couplings $h\zeta_{2R}\zeta_{2R}$ and $h\zeta_{2I}\zeta_{2I}$
will be proportional to the mass of the dark Higgses 
 and are therefore negligible for MeV dark Higgses.

The decay rate for $h \to \zeta_a \zeta_b$ can be found in Appendix A.3.
In our case only $h \to \zeta_{iR} \zeta_{jR}$ and  
$h \to \zeta_{iI} \zeta_{jI}$ exists. Furthermore, provided that the alignment condition of 
${\cal M}^2_{R,I}= v {\cal F}$ can be satisfied, only $h \to \zeta_{2R} \zeta_{2R}$
and $h \to \zeta_{2I} \zeta_{2I}$ will contribute to the SM Higgs invisible width.
The other diagonal decays $h \to \zeta_{1R} \zeta_{1R}$
and $h \to \zeta_{1I} \zeta_{1I}$ will be kinematically not accessible if 
we assume $m_{\zeta_{1R}}$ and $m_{\zeta_{1I}}$ are larger than $m_h/2$.

%%%%%%%%%%%%%%%%%%%%%%%%%%%%%%%%%%%
\subsubsection{Z Decay Width}

The measurement of $Z$-boson total width $\Gamma_Z$ at LEP 
set stringent bounds on 
any extra contribution  $\Delta \Gamma_Z$ from new decay channels.
In our case, $Z$ can decay to $\zeta_{2R} \zeta_{2I}$ through the mixing of
the neutral component of inert doublet with the complex singlet.

The $Z\zeta_a\zeta_b$ couplings are listed in Table 1 in Appendix A.3
and the corresponding tree-level decay width for each channel 
is given by Eq.~(\ref{Z2zetaazetab}) in Appendix A.4.
We will only consider the decay mode $Z \to \zeta_{2R} \zeta_{2I}$  since
other modes are presumably kinematically forbidden.
Ignoring the masses in the final state, we have
\begin{equation}
\Gamma_{Z \to \zeta_{2R} \zeta_{2I} }\approx 
\frac{ \sin^2 \theta_R \sin^2 \theta_I \sqrt{2} G_F m_Z^3}{48 \pi}  \; .
\label{eq:Zwidth}
\end{equation}
From the quoted LEP value $\Gamma_Z= 2.4952 \pm 0.0023$ GeV 
and the SM prediction $\Gamma_Z^{\rm SM} = 2.4961 \pm 0.0010$ GeV  
\cite{Beringer:1900zz},
one can estimate the maximum allowed non-standard contribution to 
$\Delta \Gamma_Z^{\rm max}$ is about 4.2 MeV at 95\% C.L.
Requiring that $\Gamma_{Z \to \zeta_{2R} \zeta_{2I} }\leq 4.2$ MeV, 
one can set the following limits on the mixing angle:
\beq
\sin\theta_R \sin \theta_I & \leq & 0.23 \; ,\\
\sin\theta_R\approx \sin \theta_I & \leq & 0.47 \; .
\label{eq:mix-lim}
\eeq

%%%%%%%%%%%%%%%%%%%%%%%%%%%%%%%%%%%
\subsubsection{S and T Parameters}

 If the scale of new physics is much larger 
than the electroweak scale, virtual effect of the new 
particles in the loops are expected to contribute  through 
vacuum polarization corrections to the electroweak precision 
observables. These corrections are known as oblique corrections 
and are parameterized by $S$, $T$ and $U$ parameters \cite{Peskin:1991sw}.
In our case, the inert Higgs doublet couples to the $W$ and $Z$ gauge bosons via the 
covariant derivative. 
Due to mixing effects, the complex singlet $\chi$ will couple to the
weak gauge bosons as well.
Therefore, both $\eta$ and $\chi$  will contribute to $S$ and $T$ parameters 
which are very well constrained by electroweak precision data.
Analytic formulas for $\Delta S$ and $\Delta T$ modified by the mixing angles 
as compared with the IHDM formulas are collected in Appendix A.5 for convenience.
Thus our model will remain viable as long as $\Delta S$ and $\Delta T$ are 
compatible with the fitted values  \cite{Baak:2014ora} which are given by:
\begin{eqnarray}
 \Delta S = 0.06\pm0.09 \quad\quad {\rm and} \quad\quad
 \Delta T = 0.10\pm0.07 \; .
\label{eq:ST}
\end{eqnarray}
%

%%%%%%%%%%%%%%%%%%%%%%%%%%%%%%%%%%%
\subsubsection{LEP Limits}

Due to the $Z_2$ symmetry, all interactions that involve the dark Higgses must
contain a pair of them. 
The precise measurements of 
$W$ and $Z$ widths at LEP \cite{Beringer:1900zz} can be used to set a limit 
on the mass of the inert
Higgses.  In order not to significantly modify the decay widths of $W$ and
$Z$,  we request that the channels $W^\pm\to \{\zeta_{iR}\eta^\pm, 
\zeta_{iI}\eta^\pm \}$ 
and/or $Z\to \{ \zeta_{iR} \zeta_{jI},\eta^+\eta^-\}$ are kinematically closed. 
This leads to the following  constraints: 
\beq
&&m_{\zeta_{iI}}+m_{\eta^\pm}>m_W \quad , \quad  m_{\zeta_{iR}}+m_{\eta^\pm}>m_W\\
&& m_{\zeta_{iR}}+ m_{\zeta_{jI}}  >m_Z \quad , \quad m_{\eta^\pm}>m_Z/2
\quad , \quad
 i,j=1,2
\eeq

At $e^+e^-$ colliders, the production mechanism for inert Higgs is
\beq
e^+e^- \to \eta^\pm \eta^\mp \quad , \quad e^+e^- \to \zeta_{iR} \zeta_{jI} \; ,
\eeq
while at hadron machines we have
\beq
&&q\bar{q} \to \eta^\pm \eta^\mp \quad , \quad q\bar{q} \to \zeta_{iR}
\zeta_{jI} \; , \\
&&q'\bar{q} \to \eta^\pm \zeta_{iR} \quad , \quad q'\bar{q} \to \eta^\pm
\zeta_{iI} \; .
\eeq
Because of $Z_2$, the inert Higgs can not decay to SM fermions.
Thus the LEPII and Tevatron searches for charged Higgs and neutral Higgs 
 can not be applied to our model. The inert charged Higgs can decay via
$\eta^\pm \to W^\pm\zeta_{2R}, W^\pm\zeta_{2I}$ 
or through cascade decay via $\eta^\pm \to W^\pm 
\zeta_{1R}\to W^\pm Z \zeta_{2I}$ or  $\eta^\pm \to W^\pm 
\zeta_{1I}\to W^\pm Z \zeta_{2R}$. 
Similarly, the neutral dark Higgses $\zeta_{1R}$ and $\zeta_{1I}$ can decay into $Z \zeta_{2I}$ and 
$Z \zeta_{2R}$ respectively,
or through cascade decays like 
$\zeta_{1R}\to \eta^\pm W^\mp \to W^\pm W^\mp \zeta_{2I}$ and 
$\zeta_{1I}\to \eta^\pm W^\mp \to W^\pm W^\mp \zeta_{2R}$.
In all cases the final states
of the these production mechanisms both at lepton or hadron colliders
would be multi-leptons or  multi-jets, depending on the 
decay products of $W^\pm$ and $Z$, plus missing energies carried by the
dark Higgses.

To certain extent, the signatures for the inert charged or neutral 
Higgses would be similar to the supersymmetry searches for charginos and 
neutralinos at the $e^+e^-$ or hadron colliders. Detailed phenomenological 
implications of this model at the LHC are interesting to explore 
but it is beyond the scope of this present work.

%%%%%%%%%%%%%%%%%%%%%%%%%%%%%%%%%%%
\subsubsection{Constraints from LHC}\label{subsubLHC}

Both ATLAS and CMS experiments of the 
LHC run at $7\oplus 8$ TeV confirmed the discovery of a scalar particle with mass around 125 GeV identified to be the Higgs 
field $h$ in our model. 
Both groups performed several measurements on this scalar particle couplings to the SM particles such as $W^+W^-$,
$ZZ$, $\gamma\gamma$ and $\tau^+ \tau^-$ with $20-30$\% uncertainties,
while for $b\bar{b}$ it suffers from larger uncertainty of
$40-50$\%. 
Recently ATLAS~\cite{Aad:2015gba} published an updated analysis of $7\oplus 8$ TeV
data in which the signal strengths $2.7^{+4.6}_{-4.5}$ for $h\to \gamma Z$ and 
$-0.7^{+3.7}_{-3.7}$ for $h\to \mu^+\mu^-$ were reported.
Basically, all the LHC data collected so far indicates that the 125 GeV boson couplings to the SM particles are very much SM-like.
One of the main tasks of the new LHC run at 13 TeV would be to
improve all the aforementioned measurements and probe for new ones, 
such as $h\to \gamma Z, \, \mu^+\mu^-$ and perhaps the trilinear self-coupling of the Higgs.
It is expected that the new run of LHC will narrow down the 
uncertainties of $h b \bar b$ and $h \tau^+ \tau^-$ measurements  
to $10-13\%$ and $6-8\%$  respectively. 
In the future, if the high luminosity option for LHC  (HL-LHC) is available,
it can do much to ameliorate the uncertainties to $4-7\%$ ($h b \bar b$) 
and $2-5\%$ ($h \tau^+ \tau^-$)~\cite{accuracy1}; while
for the $e^+e^-$ Linear Collider (LC), these uncertainties can be cut down further to 
 $0.6\%$ ($h b \bar b$) and $1.3\%$ ($h \tau^+ \tau^-$)~\cite{accuracy1,accuracy2}.

While the tree level SM Higgs couplings to fermions and to weak gauge bosons
in our model are identical to the SM one, the loop mediated processes 
such as $h\to \gamma\gamma$ and  $h\to \gamma Z$ will receive additional 
contributions from inert charged Higgs loop that can either enhance or suppress their 
partial widths \cite{Arhrib:2012ia}.
On the other hand, the invisible decay of the SM Higgs into dark Higgs pair 
is very much suppressed in our model. 
As a consequence the total width of the SM Higgs will be modified slightly
through the additional charged Higgs contributions 
in the $h\to \gamma\gamma$ and $h\to \gamma Z$ modes.
ATLAS and CMS collaborations usually present their results 
in terms of the so-called signal strengths.
For a given production channel and a given decay mode of the SM Higgs, 
the signal strength is defined as
\begin{eqnarray}
R_{YZ} \equiv \frac{\sigma(h+X) \times {\rm Br}(h\to YZ) }{\sigma^{\rm SM}(h+X) 
\times {\rm Br}^{\rm SM}(h\to YZ) } \; ,
\end{eqnarray}
where the Higgs mass is evaluated to be the same in both numerator and denominator.

In our analysis for the signal strengths, we will use the following 
ATLAS results \cite{Aad:2015gba}:
\begin{itemize}
\item $h\to \gamma\gamma$: $R_{\gamma\gamma}=1.17\pm 0.27 $
\item $h\to ZZ$: $R_{ZZ}=1.44^{+0.40}_{-0.33} $
\item $h\to W^+W^-$: $R_{WW}=1.16^{+0.24}_{-0.21}  $
\item $h\to \tau^+\tau^-$: $R_{\tau^+\tau^-}=1.43^{+0.43}_{-0.37}  $
%\item $h\to \gamma Z$ , $h\to b\bar{b}$ and $h\to \mu^+\mu^-$ are not included
%  since their errors are still very large at the time being.
\end{itemize}
%

%%%%%%%%%%%%%%%%%%%%%%%%%%%%%%%%%%%
\section{Numerical Results}

We now present our numerical results with the implementation 
of all the theoretical and experimental constraints on the parameter space 
discussed in the previous Section.
Let us classify the dimensionless
parameters $\lambda_i$ in the scalar potential 
into two different sets according to the following two types of constraints:
\begin{itemize}
\item 
First set of constraints includes
the unitarity constraints in Eq.~(\ref{constraint}), vacuum stability 
constraints in Eqs.~(\ref{eq:lam67positif})-(\ref{eq:vac2}) and also non-tachyonic masses in
Eqs.~(\ref{eq:tach1})-(\ref{eq:tach3}). We refer this set of constraints as $C_1$.
\item 
The second set of constraints contains 
the invisible decay of the $Z$ boson in Eq.~(\ref{eq:mix-lim}), $\Delta S$ and $\Delta T$ 
constraints in Eq.~(\ref{eq:ST}), signal strength constraints on $R_{\gamma\gamma}$, 
$R_{WW}$, $R_{ZZ}$ and $R_{\tau^+\tau^-}$ listed in the end of Section~(\ref{subsubLHC}). 
We also require the masses of 
$\zeta_{1R}$, $\zeta_{1I}$ and $\eta^\pm$ to be heavier than 100 GeV. 
We refer this set of constraints as $C_2$.
\end{itemize}
Since $\lambda_1$ is fixed by the SM Higgs mass, we scan over the other 
$\lambda_i \in {\cal P}$ in the following range
\begin{eqnarray}
&& 0< \lam_{2,3}\leq 4\pi \; ,\\
&& |\lam_{4,5,6,7}|\leq 4\pi \; .
\label{eq:scanla}
\end{eqnarray}

For the dimensional mass parameters in the scalar potential, 
$m_1^2$ is fixed by the SM Higgs mass, and
$m_2^2$ and $m_3^2$ are fixed by Eq.~(\ref{eq:tunem23})
in order to suppress the invisible decay of the SM Higgs.
For the  $m_4^2$ and $\mu$ parameters, they will chosen in such a way to allow for MeV dark Higgses 
$\zeta_{2R}$ and $\zeta_{2I}$. 
Recall that their masses are provided by the smaller eigenvalues of the two 
mass matrices in Eq.~(\ref{eq:mix}), 
\beq
m_{\zeta_{2R},\zeta_{2I}}^2=\frac{1}{2} \left( A+C- \sqrt{(A-C)^2+4 B^2} \right) \; ,
\label{eq:masses}
\eeq
where $A$, $B$ and $C$ are given by
~\footnote{For the heavier states $\zeta_{1R}$ and $\zeta_{1I}$, their masses are given by 
$\left( A+C+ \sqrt{(A-C)^2+4 B^2} \right)/2$.
}
\beq
\label{A}
A &= & m_2^2 + \frac{1}{2} \left( \lambda_4 + \lambda_5 \right) v^2 \; ,\\
\label{B}
B & = & \frac{1}{\sqrt 2}\, \mu v\; ,\\
\label{C}
C & = & m_3^2 + \frac{1}{2} \lambda_6 v^2 \pm m_4^2 \; .
\eea
To obtain very light dark Higgses, we fine tune
$A+C$ and $\sqrt{(A-C)^2+4 B^2}$ to be almost the same size. Define
\beq
\epsilon \equiv\frac{1}{2} (A+C - \sqrt{(A-C)^2+4 B^2}) \; .
\eeq
Assuming $\epsilon$ is small and dropping the $\epsilon^2$ term, we have
\beq
\label{eq:b2sq}
B^2=\mu^2 v^2/2\sim AC -\epsilon (A+C) \; .
\eeq
For given $\epsilon$, $A$ and $C$, the $\mu$ parameter is determined. 
If we further drop the $\epsilon$ term in Eq.~(\ref{eq:b2sq}), it would 
give an additional constraint on the sign of the product $AC>0$:
\beq
AC= (m_2^2 + (\lambda_4 + \lambda_5) v^2/2)(m_3^2 + \lambda_6 v^2/2)
=(\lambda_4 + \lambda_5) (\lambda_6) v^4 >0 \; ,
\label{AtimesC}
\eeq
where Eq.~(\ref{eq:tunem23}) has been applied in the last equality.
Thus $\lambda_4 + \lambda_5$ and  $\lambda_6$ should have the same sign. 
On the other hand, neglecting $m^2_4$ in $C$, 
Eqs.~(\ref{eq:tunem23}), (\ref{A}) and (\ref{C}) imply
\beq
A+C=(\lambda_4 + \lambda_5 + \lambda_6) v^2> 0 \; .
\label{AplusC}
\eeq
Combining Eqs.~(\ref{AtimesC}) and (\ref{AplusC}), one can conclude 
$\lambda_4 +\lambda_5$ and  $\lambda_6$ should be positive, at least for small $\epsilon$ and $m_4^2$.
Numerically, we set $\epsilon/{\rm GeV}^2 = 10^{-4}$ and $m^2_4/{\rm GeV}^2 = 10^{-5}$ in our analysis.

\begin{figure}[hptb]
\includegraphics[width=3.2in]{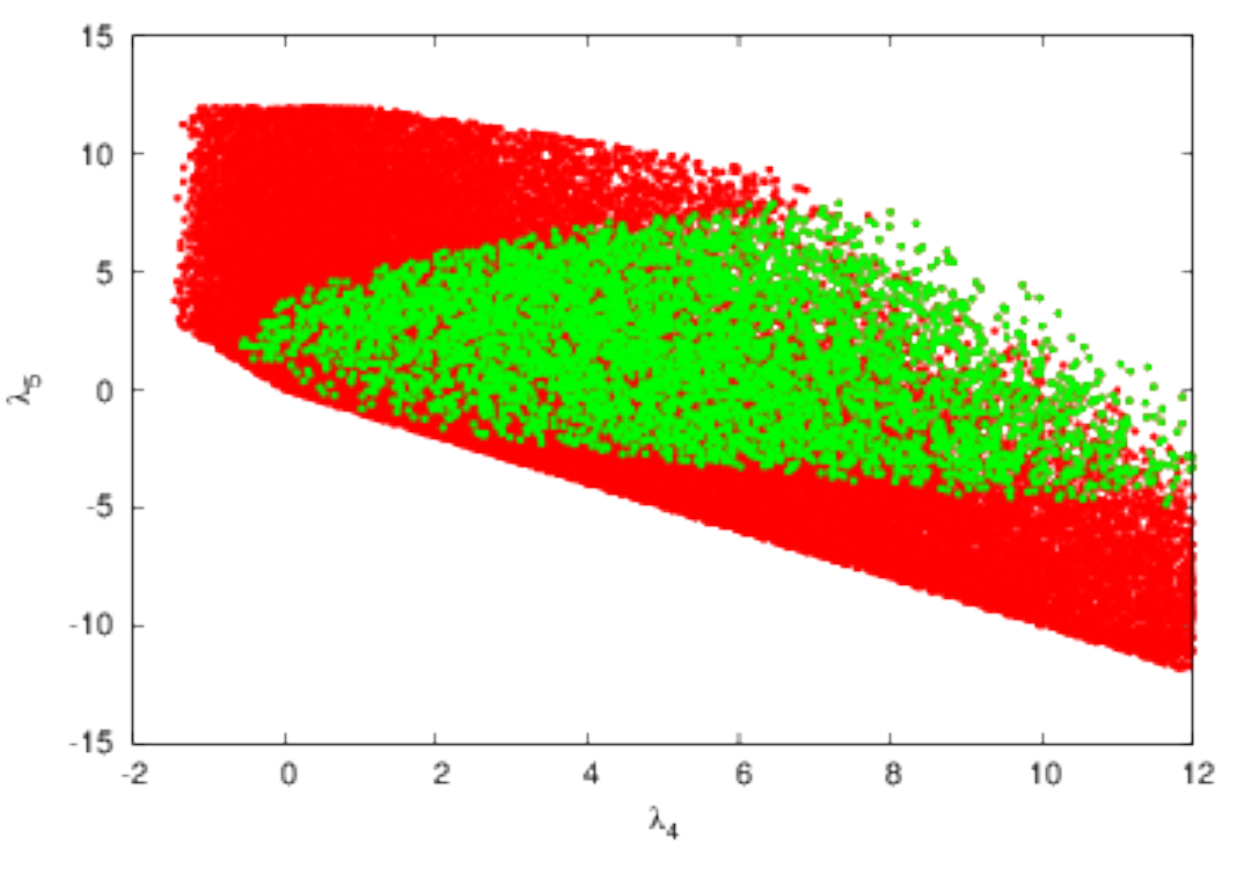}
\includegraphics[width=3.2in]{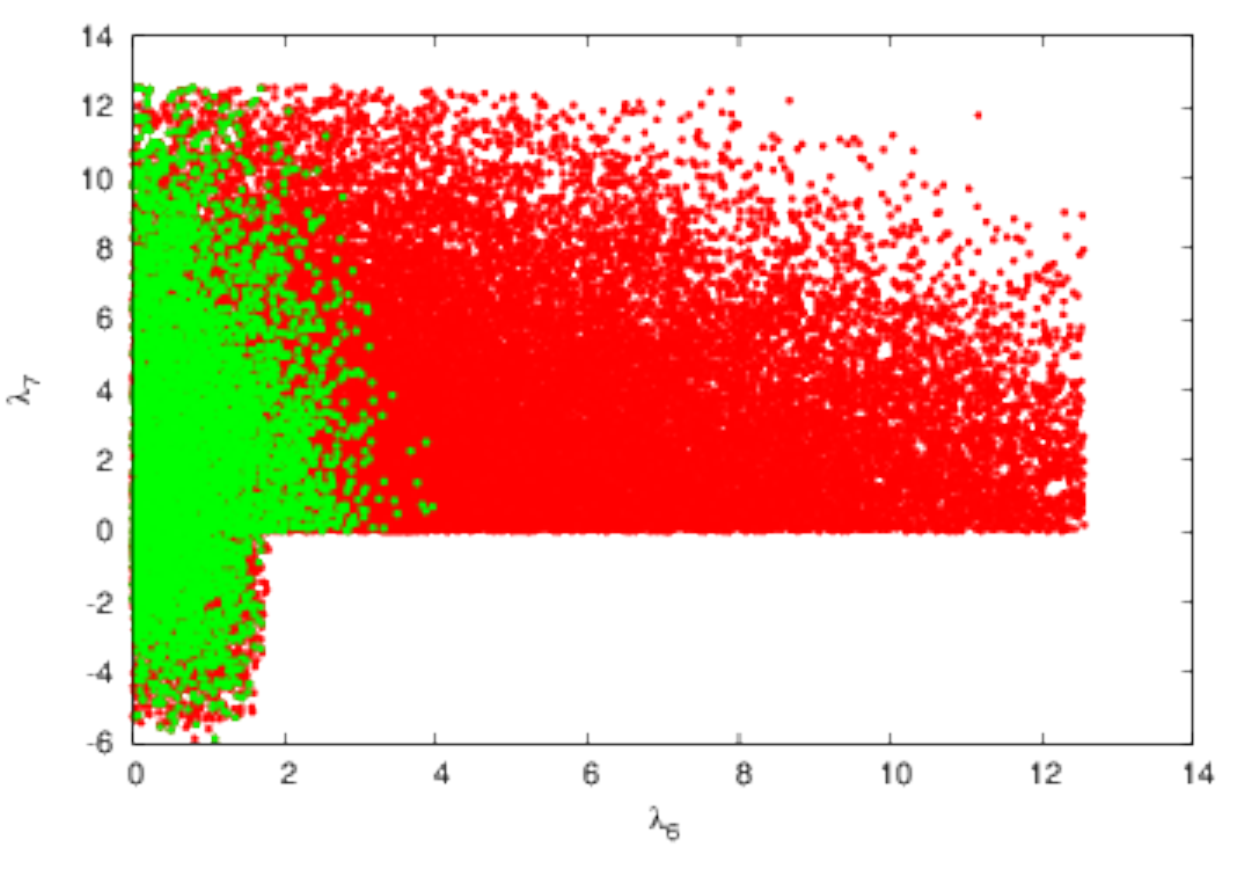}
\caption{
Allowed range for $(\lam_4,\lam_5)$ (left) and $(\lam_6,\lam_7)$ (right).
Red points pass $C_1$ set, green points pass both $C_1$ and $C_2$ sets. }
\label{fig:scanlam}
\end{figure}

A systematic scan on $\lambda_i$ in the range defined in Eq.~(\ref{eq:scanla}) indicated that 
$\lam_2$ and $\lam_3$ are not very much restricted by all the above
constraints. In Fig.~(\ref{fig:scanlam}) we illustrate the allowed range for 
$(\lam_4,\lam_5)$ (left panel) and for $(\lam_6,\lam_7)$ (right panel).
Red points pass $C_1$ set of constraints while green points pass both $C_1$
and $C_2$. 

We have checked that all the red points in the left panel of Fig.~(\ref{fig:scanlam})
fall in the following domain:
\beq
 |\lam_4+2 \lam_5|\leq 8 \pi \ \ \ {\rm and}  \ \ \ |\lam_4-\lam_5|\leq 8 \pi \; ,
\eeq
which are the unitarity constraints. 
Imposing the vacuum stability constraints reduce further the above domain. 
As one can see from the plot in the left panel, imposing merely the constraint set $C_1$,
$\lam_5$ could be either positive or negative while
$\lam_4$ is mostly positive except for a small negative range of $[-1.3,0]$. 
This small negative range for $\lam_4$ is further reduced 
when we apply the $C_2$ constraint set. 
Later we will see that the sign of $\lam_4$ is important for 
charged Higgs contribution to $h\to \gamma\gamma$.

From our previous discussion we demonstrated that under our assumptions $\lambda_6$ is positive. 
It is clear from the plot at the right panel of Fig.~(\ref{fig:scanlam}) 
that when $\lam_6$ and $\lambda_7$ are 
both positive, the $C_1$ constraint set does not restrain $\lam_6$ and $\lam_7$ 
too much. 
Even when both $C_1$ and $C_2$ are imposed, 
$\lam_7$ is not very much constrained while the range of $\lam_6$  has shrunk significantly.
This is due to the fact that $\lam_7$ does not contribute to the masses of dark Higgses 
while $\lam_6$ does. 

\begin{figure}[hptb]
\includegraphics[width=3.2in]{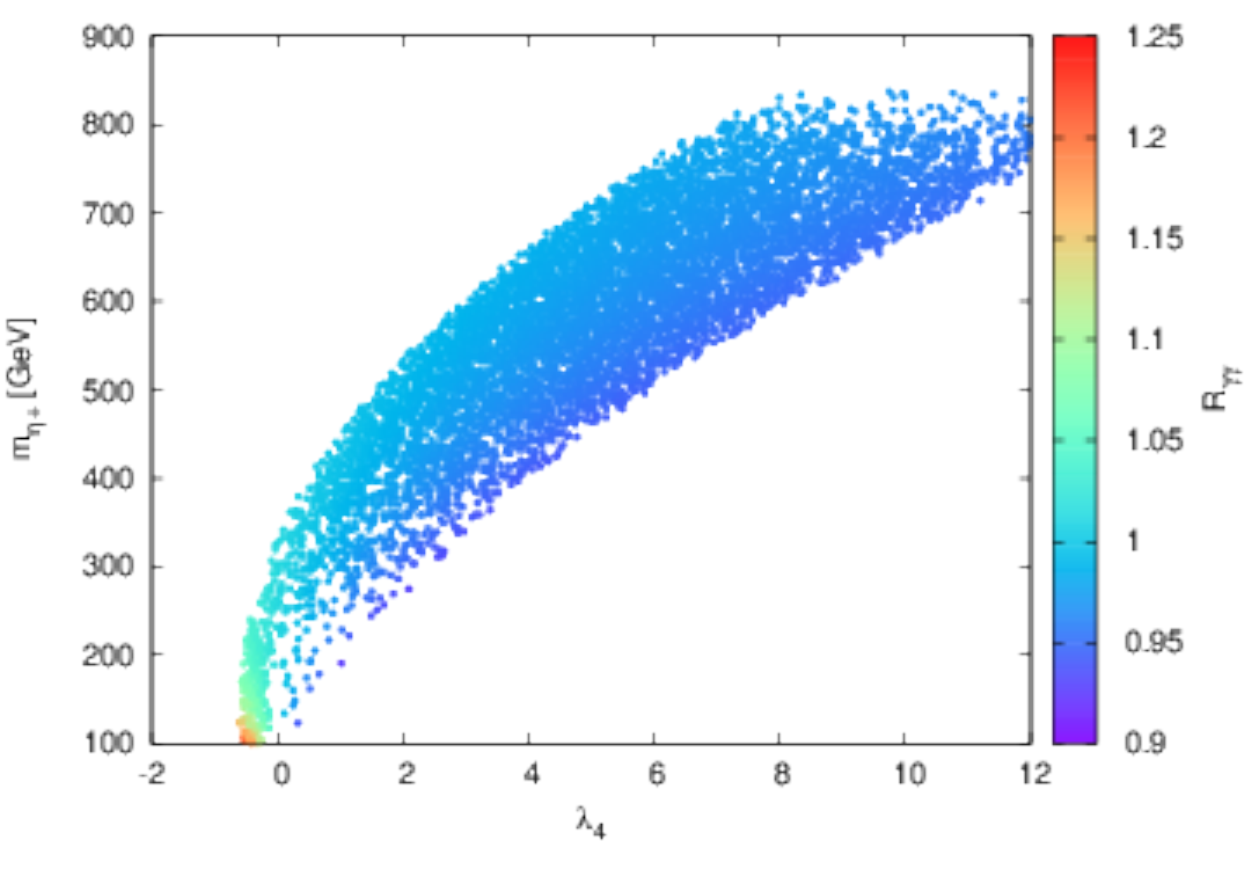}
\includegraphics[width=3.2in]{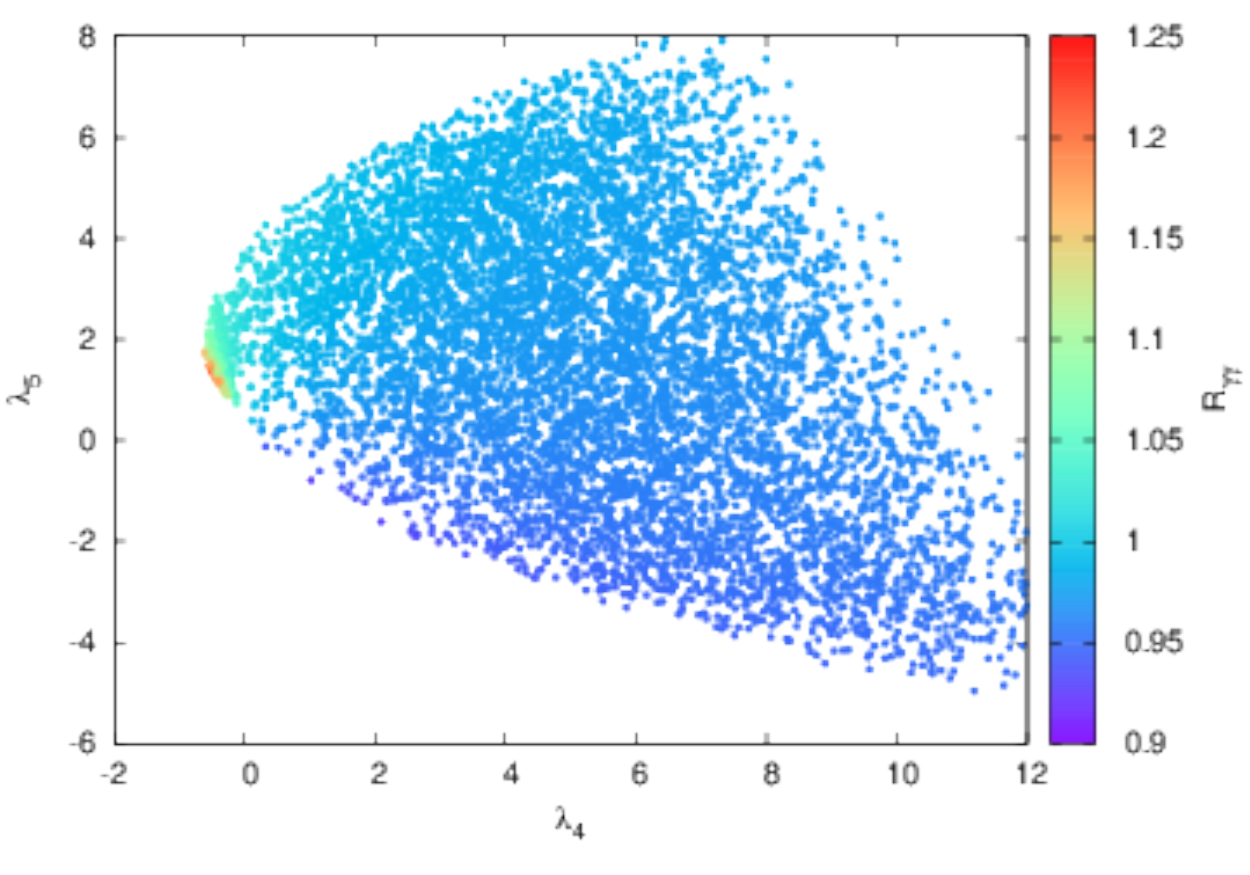}
\caption{
Scatter plot for $R_{\gamma\gamma}$ in  
$(\lam_4,m_{\eta^\pm})$ plane (left)  and in
  $(\lam_4,\lam_5)$ plane (right). All points pass both $C_1$ and $C_2$ sets. }
\label{fig:rgaga}
\end{figure}

In the left and right panels of Fig.~(\ref{fig:rgaga}) we present the scatter plots of the signal strength 
$R_{\gamma\gamma}$, represented by the color palettes located at the right sides of both panels,
on the $(\lam_4,m_{\eta^\pm})$ and $(\lam_4,\lam_5)$ planes respectively. 
In these two plots, both $C_1$ and $C_2$ constraint sets are imposed.
In our model, since the SM Higgs is produced exactly the same way as in the SM, the production
cross sections in the numerator and denominator of $R_{\gamma\gamma}$ cancel, and
the signal strength is simply given by the ratio of branching fractions. 
Thus $R_{\gamma\gamma}$ is independent of the LHC energy at Run 1 or 2. 

As is well known the loop contributions in $h\to \gamma\gamma$ is fully dominated by $W^\pm$ with some
subleading contribution from top quark which interferes destructively 
with the $W^\pm$. 
As alluded earlier, $h\to \gamma\gamma$ receives additional contribution from 
charged Higgs $\eta^\pm$ in this model \cite{Arhrib:2012ia}. The coupling of the SM Higgs to the $\eta^\pm$ 
pair is proportional to $\lam_4$.
If $\lam_4$ is negative (positive) then the $\eta^\pm$ loop is
constructively (destructively) interference with the $W^\pm$s, resulting in 
an enhanced (suppressed) $h\to \gamma\gamma$ rate with
respect to SM one. By comparing with the color palettes for $R_{\gamma\gamma}$ 
on the right side of both panels of Fig.~(\ref{fig:rgaga}), it is evident that $R_{\gamma\gamma}$ 
is enhanced for negative $\lambda_4$ but suppressed for positive $\lambda_4$. 
Note that $\lambda_4$ is restricted only to a small range of negative $\lambda_4$
$[-0.65,0]$ which could enhance $h\to \gamma\gamma$ rate 
 with respect to SM. This range of negative $\lam_4$  
corresponds to $\lambda_5$ in the range $[0.5, 3.7]$.
These two ranges for $\lam_4$ and $\lam_5$ imply that 
the charged Higgs $\eta^\pm$ is in $[100,325]$ GeV range where 
 $R_{\gamma\gamma}>1$.  
It is clear from the left panel of Fig.~(\ref{fig:rgaga}) that larger $\eta^\pm$ mass (and so as the two other neutral dark Higgses $\zeta_{1R,1I}$) say 500 GeV is also possible. But then the signal strength of $R_{\gamma\gamma}$ would be very close to its SM value.

\begin{figure}[hptb]
\begin{center}
\includegraphics[width=3.2in]{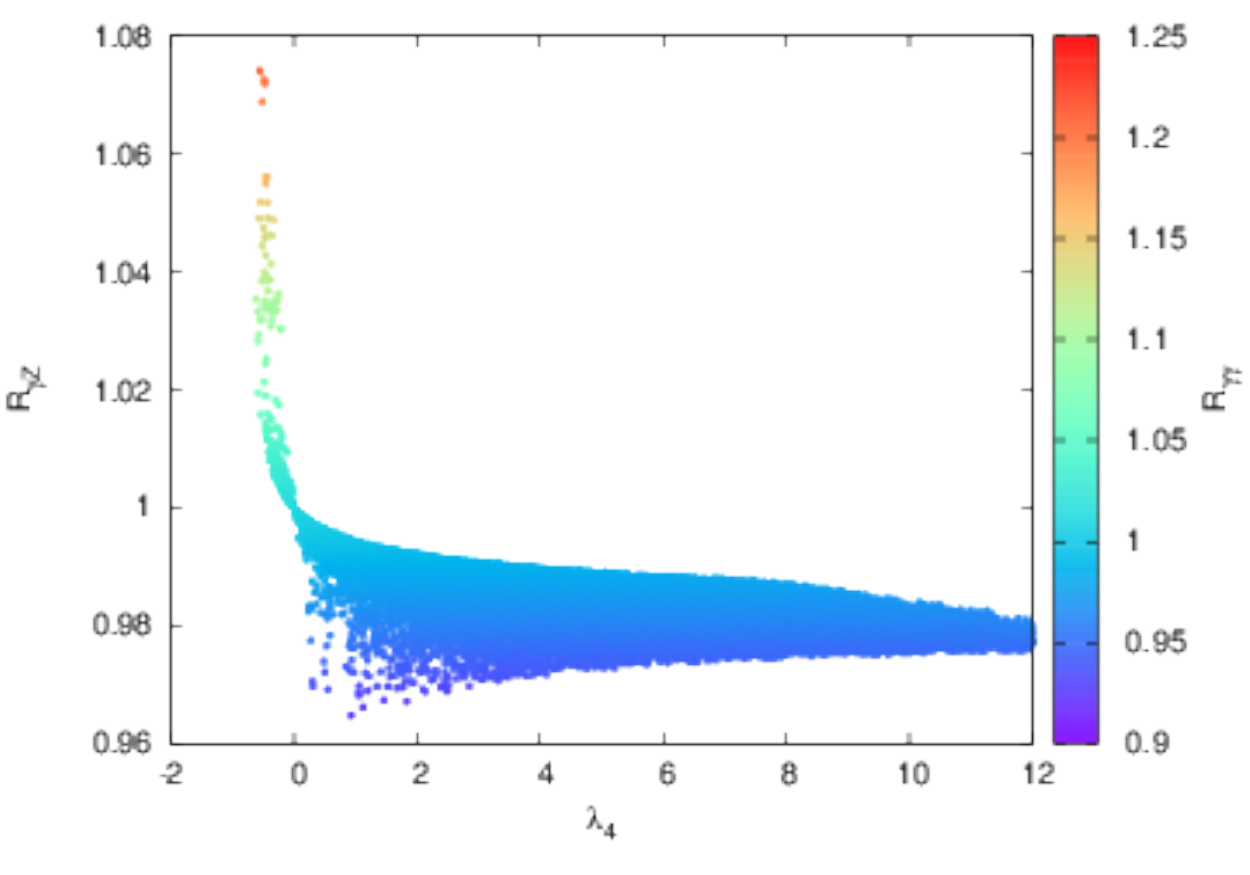}
\includegraphics[width=3.2in]{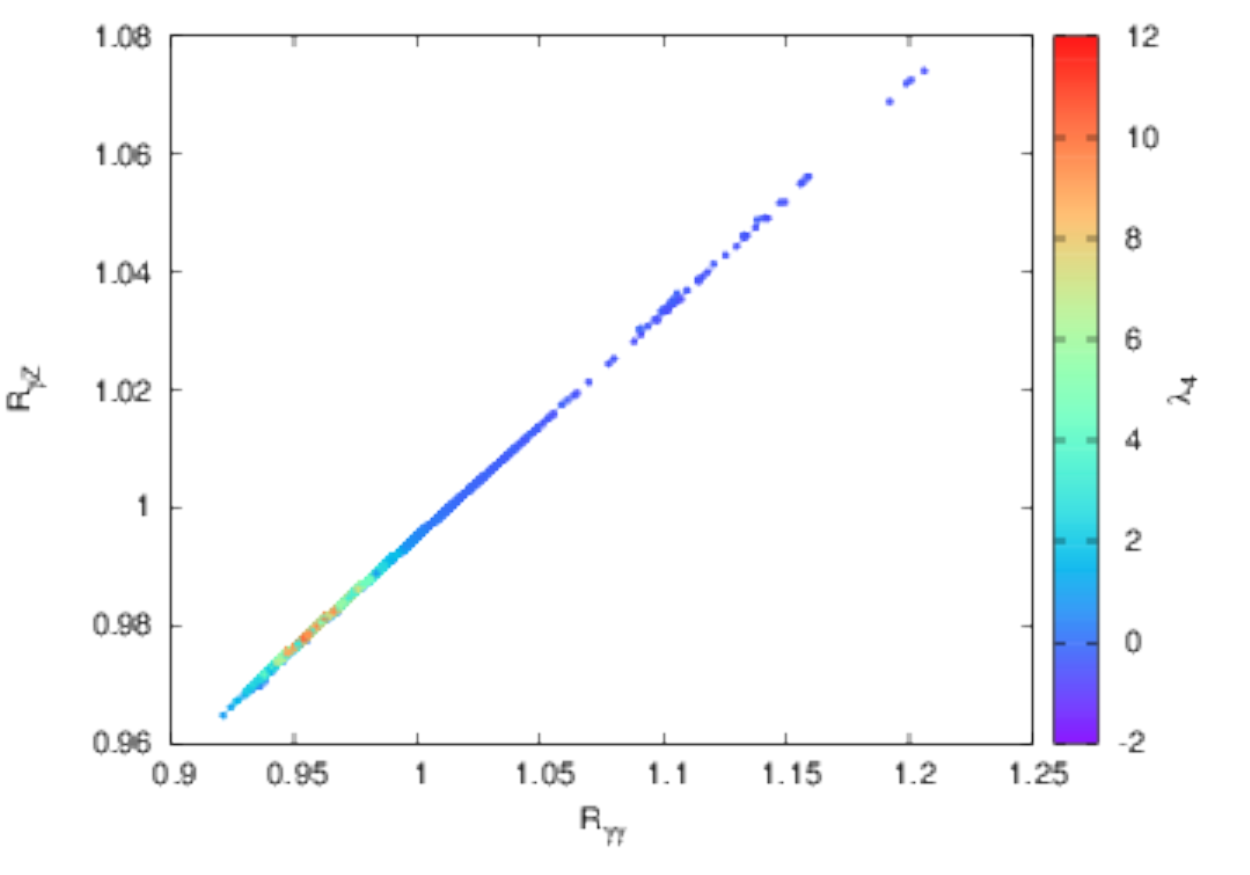}
\end{center}
\caption{
(Left) $R_{\gamma Z}$ as a function of 
$\lam_4$ with $R_{\gamma \gamma}$ shown in palette at the right.
(Right) Correlation between $R_{\gamma Z}$ and $R_{\gamma\gamma}$ with
$\lambda_4$ shown in palette at the right.}
\label{fig:rgaZ}
\end{figure}

In Fig.~(\ref{fig:rgaZ}) we illustrate $R_{\gamma Z}$ and its correlation with 
$R_{\gamma\gamma}$. In the left panel, we show $R_{\gamma Z}$ as a function of 
$\lam_4$ while scanning all other parameters. As in the $R_{\gamma\gamma}$ case,
$R_{\gamma Z}$ is enhanced for negative $\lam_4$ but suppressed for positive $\lam_4$ 
with respect to SM. In the right panel we show the correlation between 
$R_{\gamma\gamma}$ and $R_{\gamma Z}$. At the point $\lam_4=0$, 
the charged Higgs contribution vanishes and both $R_{\gamma\gamma}$ 
and $R_{\gamma Z}$ reduces to their SM values.
It is interesting to note that for $R_{\gamma\gamma}>1$ we have 
$1<R_{\gamma Z}<R_{\gamma\gamma}$ while for $R_{\gamma\gamma}<1$ we have 
$R_{\gamma Z}>R_{\gamma\gamma}$.
These predictions can be tested at LHC Run II.

%%%%%%%%%%%%%%%%%%%%%%%%%%%%%%%%%%%
\section{Conclusion}

In this work, we have presented a realistic renormalizable model
with one-loop induced neutrino mass via the interactions of neutrinos with 
MeV dark matter. Besides the SM doublet, one extra scalar doublet 
and one complex singlet were introduced in the scalar sector. 
Moreover, three light singlet Majorana fermions were needed for the one-loop
mechanism producing the neutrino masses. 
All these new fields transform under a global dark $U(1)_D$ symmetry,
which is broken softly into $Z_2$ by a single term in the scalar potential 
as well as by the assumed Majorana masses of the new fermion singlets.
The lightest of these Majorana fermions is the MeV warm dark matter while some of the 
new scalars mixed and can give rise to two MeV dark Higgses. 
Besides the 125 GeV SM Higgs, other heavier scalars include one charged Higgs and two neutral dark Higgses, which can have masses of  several hundreds GeV.

In order to suppress the decay of the SM Higgs into pair of dark Higgses we 
require its coupling matrix with the dark Higgses aligns with the mass matrix of 
the dark Higgses. For light dark Higgs masses, the invisible branching ratio 
of the SM Higgs into dark Higgses will then be suppressed and easily satisfies the 
global fit results as well as the LHC limits of the SM Higgs invisible width.

We have studied the theoretical as well as experimental constraints imposed on the scalar 
sector of the model in some detail. We have pinned down the parameter space of the model
consistent with these constraints. Our numerical results indicate that the proposed model 
is realistic. It is possible to accommodate both sub-eV neutrino masses and MeV dark matter in 
a renormalizable model with a global dark $U(1)_D$ symmetry softly breaking into $Z_2$. Some additional scalar particles of electroweak scale can be obtained.
Further collider implications of the model may be worthy of further investigation.

%%%%%%%%%%%%%%%%%%%%%%%%%%%%%%%%%%%

\noindent \underline{Acknowledgment}~:~
AA and TCY would like to thank the hospitality of Physics Division of NCTS Taiwan 
where this work was made progress. 
This work is supported by the U.~S.~Department of Energy under Grant No.~DE-SC0008541 (EM),
the Ministry of Science and Technology of Taiwan under grant number 104-2112-M-001-001-MY3 (TCY)
and the Moroccan Ministry of Higher Education and Scientific Research MESRSFC and CNRST:
``Projet dans les domaines prioritaires de la recherche scientifique et du d'eveloppement technologique": 
PPR/2015/6 (AA).

%%%%%%%%%%%%%%%%%%%%%%%%%%%%%%%%%%%

%%%%%%%%%%%%%%%%%%%%%%%%%%%%%%%%%%%
\section*{Appendix A}

%%%%%%%%%%%%%%%%%%%%%%%%%%%%%%%%%%%
\subsection*{A.1 Perturbative Unitarity Constraints}

To constrain the scalar potential parameters, 
one can demand that tree-level unitarity is preserved
in a variety of $2 \to 2$ scattering processes.

Since our model is a 2 Higgs doublet extended with a singlet field,
we can use the same procedure developed in \cite{Akeroyd:2000wc}
to derive the unitarity constraints. 
According to \cite{Akeroyd:2000wc}, one computes the $S$ matrix in the non-physical fields
basis where the computation is much easier.
The crucial point is that
the $S$ matrix expressed in terms of the physical fields
({\it i.e.} the mass eigenstate fields) can be transformed into
an $S$ matrix for the non-physical fields
by making a unitary transformation. The eigenvalues for the $S$ matrix should be unchanged under
such a unitary transformation. 

The first submatrix ${\cal M}_1$, corresponding to scatterings whose
initial and final states being one of the following combinations
$(w_1^+w_2^-$,$w_2^+w_1^-$, $\phi_R \eta_I$, $\eta_R\phi_I$, 
$\phi_I \eta_I$, $\phi_R \eta_R)$, is given by 
\begin{eqnarray}
{\cal M}_1 =\left(
\begin{array}{cccccc}
\lambda_4 + \lambda_5 & 0 & -\frac{\lambda_5}{2} & \frac{\lambda_5}{2} &\frac{\lambda_5}{2} & \frac{\lambda_5}{2} \\ 
 0 & \lambda_4 + \lambda_5 & \frac{\lambda_5}{2} & -\frac{\lambda_5}{2} & \frac{\lambda_5}{2} & \frac{\lambda_5}{2}\\
 \frac{\lambda_5}{2} & -\frac{\lambda_5}{2} & \lambda_4 + \lambda_5 & 0 & 0 & 0\\ 
 -\frac{\lambda_5}{2} & \frac{\lambda_5}{2} & 0 & \lambda_4 + \lambda_5 & 0 & 0\\ 
 \frac{\lambda_5}{2} & \frac{\lambda_5}{2} & 0 & 0 & \lambda_4 + \lambda_5 & 0\\ 
\frac{\lambda_5}{2} & \frac{\lambda_5}{2} & 0 & 0 & 0 & \lambda_4 + \lambda_5
\end{array}
\right) \; .
\end{eqnarray}
Its eigenvalues are determined as
\begin{eqnarray}
e_1& = & \lam_4 + 2 \lam_5 \; , \\
e_2& = & \lam_4 \; , \\
f_+& = & \lam_4 + 2 \lam_5 \; ,\\
f_{-} & = & e_2 \; , \\
f_1 & = & f_2 \; = \; \lambda_4 + \lambda_5 \; .
\end{eqnarray}

The second submatrix ${\cal M}_2$ 
corresponds to scattering with  initial and final
states belonged to one of the following states
$(w_1^+w_1^-$, $w_2^+w_2^-$,
$\frac{\phi_I\phi_I}{\sqrt{2}}$,
 $\frac{\eta_I\eta_I}{\sqrt{2}}$, $\frac{\phi_R\phi_R}{\sqrt{2}}$,
$\frac{\eta_R\eta_R}{\sqrt{2}}$, $\frac{\chi_R\chi_R}{\sqrt{2}}$, 
$\frac{\chi_I\chi_I}{\sqrt{2}}$), where the $\sqrt{2}$ accounts for
identical particle statistics. This matrix mixes doublet with singlet
states and is given by
\begin{eqnarray}
{\cal M}_2 =\left(
\begin{array}{cccccccc}
2\lambda_1 & \lambda_{45} & 
\frac{\lambda_1}{\sqrt{2}} & \frac{\lambda_1}{\sqrt{2}} & 
\frac{\lambda_4}{\sqrt{2}}& 
 \frac{\lambda_4}{\sqrt{2}} &  \frac{\lambda_6}{\sqrt{2}}
& \frac{\lambda_6}{\sqrt{2}} \\ 
\lambda_{45} & 2 \lambda_2& \frac{\lambda_4}{\sqrt{2}}  & 
\frac{\lambda_4}{\sqrt{2}} &
\frac{\lambda_2}{\sqrt{2}} & 
 \frac{\lambda_2}{\sqrt{2}}  &\frac{\lambda_7}{\sqrt{2}} & 
\frac{\lambda_7}{\sqrt{2}} \\ 
 \frac{\lambda_1}{\sqrt{2}}  & \frac{\lambda_4}{\sqrt{2}} & 
\frac{3\lambda_1}{2} 
& \frac{\lambda_{1}}{2} & \frac{\lambda_{45}}{2} & 
  \frac{\lambda_{45}}{2} & \frac{\lambda_{6}}{2} & \frac{\lambda_{6}}{2} \\
\frac{\lambda_1}{\sqrt{2}} & \frac{\lambda_4}{\sqrt{2}} & 
\frac{\lambda_1}{2}  & 
  \frac{3\lambda_1}{2} & \frac{\lambda_{45}}{2} & 
 \frac{\lambda_{45}}{2} &
  \frac{\lambda_6}{2} & \frac{\lambda_6}{2}  \\
 \frac{\lambda_4}{\sqrt{2}}  & \frac{\lambda_2}{\sqrt{2}} & 
\frac{\lambda_{45}}{2} & \frac{\lambda_{45}}{2} & \frac{3\lambda_2}{2} & 
 \frac{\lambda_2}{2} & \frac{\lambda_7}{2}& \frac{\lambda_7}{2} \\
\frac{\lambda_4}{\sqrt{2}}  & \frac{\lambda_2}{\sqrt{2}} & 
\frac{\lambda_{45}}{2} &
 \frac{\lambda_{45}}{2} & \frac{\lambda_2}{2}  & 
\frac{3\lambda_2}{2}  & \frac{\lambda_{7}}{2}  &
 \frac{\lambda_{7}}{2} \\
 \frac{\lambda_{6}}{\sqrt{2}} & \frac{\lambda_7}{\sqrt{2}}   & 
 \frac{\lambda_{6}}{2} & 
\frac{\lambda_{6}}{2} & \frac{\lambda_{7}}{2} & 
\frac{\lambda_{7}}{2}  & 
\frac{3\lambda_3}{2}  & \frac{\lambda_{3}}{2}  \\ 
 \frac{\lambda_6}{\sqrt{2}}    &\frac{\lambda_7}{\sqrt{2}} & 
\frac{\lambda_{6}}{2}  & \frac{\lambda_{6}}{2}  & 
  \frac{\lambda_{7}}{2}  & 
\frac{\lambda_{7}}{2}  & \frac{\lambda_{3}}{2}  & \frac{3\lambda_{3}}{2} 
\end{array}\right) \; ,
\end{eqnarray}
where $\lam_{45}=\lam_{4}+\lam_{5}$. This matrix has 8 eigenvalues. 
Five of them are 
\begin{eqnarray}
&& c_+=\lambda_1 \; ,\\
&& c_-= \lambda_2 \; , \\
&& s_1 =\lambda_3 \; , \\
&& a_{\pm}=\frac{1}{2} (\lambda_1+\lambda_2\pm 
\sqrt{(\lambda_1-\lambda_2)^2+4\lambda_5^2 }) \; .
\end{eqnarray}
The other 3 eigenvalues $b_{\pm}$ and $s_2$ 
are solutions of the following polynomial
\begin{eqnarray}
P(X)&=&2 [3 \lambda_2 \lambda_6^2 + (2 \lambda_4 + \lambda_5) (2 \lambda_4 
\lambda_3 + \lambda_5 \lambda_3 - 
     2 \lambda_6 \lambda_7) + 3 \lambda_1 (-3 \lambda_2 \lambda_3 + 
\lambda_7^2)] - \nonumber\\ &&
 [-9 \lambda_1 \lambda_2 + (2 \lambda_4 + \lambda_5)^2 - 
6 \lambda_{12} \lambda_3 + 
   2 (\lambda_6^2 + \lambda_7^2)] X - [3 \lambda_{12} + 2 
\lambda_3] X^2 + X^3
\end{eqnarray}
where $\lam_{12}=\lam_1+\lam_2$.

The third submatrix ${\cal M}_3$ expressed in the basis
($\phi_R \phi_I$, $\eta_R \eta_I$) is diagonal with $c_{\pm} = \lambda_{1,2}$ 
as eigenvalues as defined previously.  

With the two singlet components, more states such as ($\phi_R \chi_I$, $\eta_R \chi_I$),
($\phi_R \chi_R$, $\eta_R \chi_R$), 
($\phi_R \chi_{R,I}$, $\phi_I \chi_{R,I}$),
($\eta_R\chi_{R,I}$, $\eta_I\chi_{R,I}$), 
{\it etc}. can be constructed.
But their corresponding scattering matrices will be diagonal and lead 
to either $\lambda_6$ or $\lambda_7$ as eigenvalues. 
These scattering states will not lead to any nontrivial constraints among $\lambda_{i}$
since they are required to be perturbative, namely $\vert \lambda_i \vert \leq 4 \pi$ for all $i$.

In our analysis we also include the following two body scattering processes
among the 8 charged states ($\phi_R w_1^+ $,  $\eta_R w_1^+ $,
$\phi_I w_1^+ $, $\eta_I w_1^+$, $\phi_R \eta^+$, $\eta_R \eta^+$, 
$\phi_I \eta^+$,  $\eta_I \eta^+$). This submatrix only lead to one 
additional constraint which is
\beq
 p_1=\lambda_4 - \lambda_5 \; .
\eeq
The others are duplicated with the previous cases.

With the two singlet components, we can also construct charged states like $(\chi_R
w_1^+,\chi_I w_1^+)$ and $(\chi_R
\eta^+,\chi_I \eta^+)$ which decouple from the previous charged scattering
processes. Again, the scattering matrices in these cases are diagonal
with eigenvalues $\lambda_{6}$ and $\lambda_{7}$ respectively.

All the eigenvalues shown in this appendix are required to satisfy the perturbative unitarity constraints 
as given by Eq.~(\ref{constraint}).

%%%%%%%%%%%%%%%%%%%%%%%%%%%%%%%%%%%
\subsection*{A.2 Vacuum Stability Constraints on Scalar Potential}

At large field values the potential Eq.~(\ref{eq:pot})
is  dominated only by the part containing the terms that are quartic in
the fields
\begin{eqnarray}
V_{\rm quartic}
&=& \frac{1}{2} \lambda_1 (\Phi^\dagger \Phi)^2 
+ \frac{1}{2} \lambda_2 (\eta^\dagger \eta)^2 +
\lambda_4 (\eta^ \dagger \eta)(\Phi^\dagger \Phi) 
+ \lambda_5 (\eta^ \dagger \Phi)(\Phi^\dagger \eta) \nn\\
&+& 
 \frac{1}{2} \lambda_3 (\chi^* \chi)^2
+ \lambda_6 (\chi^* \chi)(\Phi^\dagger \Phi)
+ \lambda_7 (\chi^* \chi)(\eta^\dagger \eta)  \; .
\label{eq:pot4}
\end{eqnarray}
The study of $V_{\rm quartic}$ will thus be sufficient to obtain the main
constraints from vacuum stability considerations.

Following \cite{ElKaffas:2006nt}, we adopt the following
parameterization of the fields. First, we introduce 
the unit spinors $\hat{\Phi} $ and $\hat{\eta} $ such that
\beq
&&\Phi =|\Phi| \hat{\Phi} \quad , \quad \eta =|\eta| \hat{\eta} \quad , \quad 
\Phi^+\Phi= |\Phi|^2 \quad , \quad \eta^+\eta= |\eta|^2 \nn
\\
&& 
\Phi^+ \eta = |\eta|  |\Phi| (\hat{\Phi}^+ \cdot \hat{\eta}) \; .
\eeq
$(\hat{\Phi}^+ \cdot \hat{\eta})$ is a scalar product of 2 unit spinors which can
be written as $a+ib=\rho e^{i\gamma}$ ( $\rho=|a+ib| \in [0,1]$).
We then have the following parameterization
\beq
&& |\Phi| = r \cos \theta \sin \phi \;,   \\ 
&& |\eta| = r \sin \theta \sin \phi \; , \\
&& \Phi^\dagger \eta= |\Phi||\eta| \rho e^{i\gamma}=r^2 \cos\theta \sin\theta \sin^2\phi \; , \\
&& |\chi| = r \cos  \phi \; ,
\eeq
when $\Phi$, $\eta$ and $\chi$ scan all the field space, 
$r$ scans the domain $[0, \infty)$, $\rho \in [0,1]$, and
the angles $\theta,\phi \in [0, \pi/2]$. The phase $\gamma$ 
will not have any effect here. Our potential does not have 
$(\Phi^\dagger \eta)^2$ as a quartic term in the potential because of dark 
$U(1)_D$ invariance.

One can rewrite the quartic terms using the new parameterization as
\beq
V_{\rm quartic} &=&r^4\{ [
\frac{\lam_1}{2}\cos^4 \theta+\frac{\lam_2}{2}\sin^4\theta + 
( \lambda_4  + \lam_5 \rho^2) \sin^2\theta \cos^2\theta ]\sin^4\phi +
\nn \\
& &\frac{\lam_3}{2}\cos^4\phi +
 [\lam_{6}\cos^2\theta +\lam_{7} \sin^2\theta] \cos^2\phi \sin^2\phi\} \; , \\
 &=&r^4 \{(
\frac{\lam_1}{2}\cos^4 \theta+\frac{\lam_2}{2}\sin^4\theta + 
(\lambda_4  +\lam_5 \rho^2) \sin^2\theta \cos^2\theta )  x^2 +
\nn \\
& &\frac{\lam_3}{2} (1-x)^2 + (
 \lam_{6}\cos^2\theta + \lam_{7} \sin^2\theta) x (1-x) 
\} \; ,
\eeq
where we have used $x=\sin\phi$. In this form, $V_{\rm quartic}/r^4$ is a second
degree polynomial in $x\in [0,1]$. One can show that $V_{\rm quartic}/r^4$ is positive
 if and only if~\footnote{$a x^2 + b (1-x)^2 + cx(1-x)=(\sqrt{a} x
   -\sqrt{b}(1-x))^2 + (c+2 \sqrt{ab}) x(1-x)$ is positive if and only if
$a>0$, $b>0$ and $c>-2\sqrt{ab}$.
}
\beq
\label{bfbA}
&& {\cal A} \equiv 
\frac{\lam_1}{2} y^2+\frac{\lam_2}{2}(1-y)^2+(\lam_4  +
\lam_5 \rho^2) y (1-y)   > 0 \; , \\
\label{bfbB}
&& {\cal B} \equiv \frac{1}{2}\lam_{3} > 0 \; , \\
\label{bfbC}
&& {\cal C} \equiv \left( \lam_{6}y+\lam_{7}(1-y) \right) > -2\sqrt{{\cal A}\,{\cal B}} \; ,
\eeq
where we used 
$y=\cos^2\theta$. The first condition (Eq.~(\ref{bfbA})) is nothing but a scalar potential 
 without the singlet field. This condition will give us the boundedness from below
 for 2 Higgs doublet model. We note that ${\cal A}$ is  a second degree polynomial in $y$.
It is positive, if and only if
\beq
&& \lam_1 >0 \quad , \quad \lam_2>0 \; , \\
&& \lam_4 + \lam_5 \rho^2 >  - \sqrt{\lam_1\lam_2} \quad , \quad \rho\in [0,1] \; .
\eeq
The last condition of the above equation gives the following 2 conditions
\beq
\lam_4 + \sqrt{\lam_1\lam_2} >0 \quad {\rm and}  \quad  
\lam_4 + \lam_5 + \sqrt{\lam_1\lam_2} >0 \; .
\eeq

With the presence of the singlet field, we have from Eqs.~(\ref{bfbB}) and (\ref{bfbC})
\begin{itemize}
\item $\lam_3>0$ from ${\cal B} >0$. 

\item 
If $\lam_6>0$ and $\lam_7 >0$, since $y\in [0,1]$ the third 
constraint $ \lam_{6}y+\lam_{7}(1-y)  > -2\sqrt{{\cal A}\,{\cal B}}$ is satisfied 
for any $\lam_6>0$ and $\lam_7 >0$. In this case they will 
be no additional constraints on $\lam_6>0$ and $\lam_7 >0$.

\item 
If $\lam_6<0$ or $\lam_7 <0$,  one has
$-2\sqrt{{\cal A}\,{\cal B}} <  \lam_{6}y+\lam_{7}(1-y) < 2\sqrt{{\cal A}\,{\cal B}}$. If not,
$\lam_{6}y+\lam_{7}(1-y) > 2\sqrt{{\cal A}\,{\cal B}}$ will lead to $\lam_{6,7}>0$
which is not the case.

Then we can rewrite the third condition 
${\cal C}>-2\sqrt{{\cal A}\,{\cal B}}$ (Eq.~(\ref{bfbC}))  as
\beq
(\lam_3 \lam_1 -\lam_6^2) y^2 +(\lam_3 \lam_2 -\lam_7^2) (1-y)^2+ 
(-2 \lam_6\lam_7 + 2\lam_3 (\lam_4 +\lam_5 \rho^2 )) y (1-y) >0 \; ,
\eeq
which is positive, if and only if
\beq
\label{eq:bfbsinglet1}
&&(\lam_3 \lam_1 -\lam_6^2) >0 \; , \\
\label{eq:bfbsinglet2}
&& (\lam_3 \lam_2 -\lam_7^2) >0 \; , \\
\label{eq:bfbsinglet3}
&& (-2 \lam_6\lam_7 + 2\lam_3 (\lam_4 +\lam_5 \rho^2 ))  >
-\sqrt{4(\lam_3 \lam_1 -\lam_6^2)(\lam_3 \lam_2 -\lam_7^2)} \; .
\eeq
\end{itemize}
The 2 constraints of Eqs.~(\ref{eq:bfbsinglet1}) and (\ref{eq:bfbsinglet2}) will give
\beq
&&\sqrt{\lam_3 \lam_1} +\lam_6 >0  \quad {\rm and}  \quad 
\sqrt{\lam_3 \lam_1} -\lam_6>0 \; , \\
&& \sqrt{\lam_3 \lam_2} +\lam_7^2 > 0  \quad {\rm and} \quad 
\sqrt{\lam_3 \lam_2} -\lam_7>0 \; .
\eeq
If we work out the third condition of Eq.~(\ref{eq:bfbsinglet3}), we get
\beq
&& - \lam_6\lam_7 + \lam_3 \lam_4 >- 
\sqrt{(\lam_3\lam_1 -\lam_6^2)(\lam_3\lam_2 -\lam_7^2)} \; , \\
&& - \lam_6\lam_7 + \lam_3 (\lam_4 +\lam_5  )>- 
\sqrt{(\lam_3\lam_1 -\lam_6^2)(\lam_3\lam_2 -\lam_7^2)} \; .
\eeq

In our analysis, we impose all the constraints derived in this appendix 
for the quartic couplings $\lambda_i$.

%%%%%%%%%%%%%%%%%%%%%%%%%%%%%%%%%%%
\subsection*{A.3 Scalar Cubic Couplings of the SM Higgs}

\begin{table}[hptb]
\caption{General coupling coefficients of $h\zeta_a \zeta_b$ and $Z\zeta_a \zeta_b$ vertices.}
\begin{center}
\begin{tabular}{l|c|c}
\hline
$(a,b)$ & $g_{ab}$ & $c_{ab}$\\
\hline\hline
$(1R,1R)$ & $\left( \left( \lambda_4 + \lambda_5 \right) \cos^2 \theta_R 
+ \lambda_6 \sin^2 \theta_R \right) - \frac{\sqrt 2}{2}\frac{\mu}{v}\sin 2 \theta_R$ & 0\\ 
$(2R,2R)$ & $\left( \left( \lambda_4 + \lambda_5 \right) \sin^2 \theta_R 
+ \lambda_6 \cos^2 \theta_R \right) + \frac{\sqrt 2}{2}\frac{\mu}{v}\sin 2 \theta_R$ & 0\\ 
$(1R,2R)$ & $ \frac{1}{2} \left( \lambda_4 + \lambda_5 - \lambda_6 \right) \sin 2 \theta_R 
+ \frac{\sqrt 2}{2}\frac{\mu}{v}\cos 2 \theta_R$ & 0\\ 
\hline
$(1I,1I)$ & $\left( \left( \lambda_4 + \lambda_5 \right) \cos^2 \theta_I 
+ \lambda_6 \sin^2 \theta_I \right) - \frac{\sqrt 2}{2}\frac{\mu}{v}\sin 2 \theta_I$ & 0\\ 
$(2I,2I)$ & $\left( \left( \lambda_4 + \lambda_5 \right) \sin^2 \theta_I 
+ \lambda_6 \cos^2 \theta_I \right) + \frac{\sqrt 2}{2}\frac{\mu}{v}\sin 2 \theta_I$ & 0 \\ 
$(1I,2I)$ & $ \frac{1}{2} \left( \lambda_4 + \lambda_5 - \lambda_6 \right) \sin 2 \theta_I 
+ \frac{\sqrt 2}{2}\frac{\mu}{v}\cos 2 \theta_I$ & 0\\ 
\hline
$(1R,1I)$ & 0 & $\cos\theta_R \cos\theta_I$ \\
$(1R,2I)$ & 0 & $\cos\theta_R \sin\theta_I$ \\
$(2R,1I)$ & 0 & $\sin\theta_R \cos\theta_I$ \\
$(2R,2I)$ & 0 & $\sin\theta_R \sin\theta_I$ \\
\hline
\end{tabular}
\end{center}
\label{table1}
\end{table}

The cubic couplings for $h \zeta_a \zeta_b$ is given by 
$-i g_{ab} v$ with $g_{ab}$ defined in the second column of Table~\ref{table1}.
The decay rate for $h \to \zeta_a \zeta_b$ is given by
\begin{equation}
\Gamma \left( h \to \zeta_a \zeta_b \right) = \frac{1}{1 + \delta_{ab}} 
\frac{1}{16\pi}\frac{v^2}{m_h} \vert g_{ab} \vert^2 
\lambda^{\frac{1}{2}} \left( 1, \frac{m_a^2}{m^2_h} , \frac{m_b^2}{m^2_h} \right) \; ,
\end{equation}
where
\be
\lambda(x,y,z) = x^2 + y^2 +z^2 - 2(xy +yz + zx) \; .
\ee
The $h \eta^+\eta^-$ coupling is simply $-i \lambda_4 v$ while the SM $hhh$ self coupling is $-i \lambda_1 v$.

%%%%%%%%%%%%%%%%%%%%%%%%%%%%%%%%%%%%%%%
\subsection*{A.4 $Z\zeta_a \zeta_b$ Couplings}

From the covariant derivative $( D_\mu \eta )^\dagger (D^\mu \eta )$, we have the following derivative couplings
\begin{eqnarray}
{\mathcal L}_{\rm int} & \supset & 
i \frac{g}{2} \left[ W^{\mu +} \left( \eta^- \overleftrightarrow{\partial_\mu} \left( \eta_R + i \eta_I \right) \right)
+ W^{\mu -} \left( \left( \eta_R - i \eta_I \right)  \overleftrightarrow{\partial_\mu} \eta^+ \right)
 \right] \nonumber \\
 & + & i e \left( A^\mu +  \left( \frac{c^2_{\theta_w} - s^2_{\theta_w}}{2 s_{\theta_w} c_{\theta_w}} \right) Z^\mu \right)
\left( \eta^- \overleftrightarrow{\partial_\mu} \eta^+ \right) 
+ \frac{g}{2 c_{\theta_w}} Z^\mu \left( \eta_R \overleftrightarrow{\partial_\mu} \eta_I \right)  \; ,
\label{derivativecouplings}
\end{eqnarray}
where the fields $\eta_{R,I}$ are related to the physical fields 
$\zeta_{1R}$, $\zeta_{2R}$, $\zeta_{1I}$ and $\zeta_{2I}$ as 
$\eta_{R}  =  \cos \theta_R \zeta_{1R} + \sin \theta_R \zeta_{2R}$ and
$\eta_{I}  =  \cos \theta_I \zeta_{1I} + \sin \theta_I \zeta_{2I}$. 
From the last term in Eq.~(\ref{derivativecouplings}), 
we get the vertex for $Z (\epsilon_\mu(k)) \to \zeta_a (p) \zeta_b(p')$ 
as $+(g/2c_{\theta_w})c_{ab}(p - p')_\mu$ 
with $c_{ab}$ defined in the last column of Table~\ref{table1}.
The decay rate for $Z \to \zeta_a \zeta_b$ is given by
\be
\label{Z2zetaazetab}
\Gamma (Z \to \zeta_a \zeta_b ) = \frac{\sqrt 2}{48 \pi} G_F m^3_Z \vert c_{ab} \vert^2 \lambda^{\frac{3}{2}}
\left(1, \frac{m_a^2}{m^2_Z}, \frac{m_b^2}{m_Z^2} \right) \; .
\ee
%

%%%%%%%%%%%%%%%%%%%%%%%%%%%%%%%%%%%
\subsection*{A.5 Formulas for the $\Delta S$ and $\Delta T$}

The analytic expressions for $\Delta S$ and $\Delta T$ can be given in terms of Passarino-Veltman
functions which have been calculated using the software packages 
FormCalc~\cite{FC} and LoopTools~\cite{LT}. 
%The scalar integrals
%are computed numerically using LoopTools package. 
The SM expressions for $S$ and $T$ have been subtracted properly, 
we give hereafter only the extra contributions $\Delta S$ and $\Delta T$.  
We take as reference point the Higgs mass $m_h = 125$ GeV, 
$m_t = 173$ GeV and assume $\Delta U=0$.

We have checked both analytically and numerically that $\Delta S$ and $\Delta T$ are UV finite and also 
independent of the renormalisation scale. In terms of the Passarino-Veltman functions $A_0$ and $B_{00}$, 
they are given by
\begin{eqnarray}
\Delta S&=&\frac{1}{ \pi m_Z^2}(
2 c_W^2 s_W2 A_0[m_{\eta^\pm}^2] - 
\cos^2\theta_I  \sin^2\theta_R
B_{00}[0,m_{\zeta_{1I}}^2,m_{\zeta_{2R}}^2]
\nonumber\\ && + 
  B_{00}[ 0, m_{\eta^\pm}^2, m_{\eta^\pm}^2] (1 - 2 s_W^2)^2 - 
 \cos^2\theta_I  \cos^2\theta_R B_{00}[0, m_{\zeta_{1R}}^2,
  m_{\zeta_{1I}}^2] 
\nonumber\\ && - 
\cos^2\theta_R  \sin^2\theta_I B_{00}[0, m_{\zeta_{1R}}^2, m_{\zeta_{2I}}^2] - 
\sin^2\theta_R  \sin^2\theta_I B_{00}[0, m_{\zeta_{2R}}^2,
  m_{\zeta_{2I}}^2]\nonumber\\ && - 
  B_{00}[m_Z^2, m_{\eta^\pm}^2, m_{\eta^\pm}^2] + 
\cos^2\theta_I\sin^2\theta_R B_{00}[m_Z^2, m_{\zeta_{1I}}^2,
  m_{\zeta_{2R}}^2]
\nonumber\\ && + 
\cos^2\theta_I  \cos^2\theta_R B_{00}[m_Z^2, m_{\zeta_{1R}}^2, m_{\zeta_{1I}}^2] + 
  \sin^2\theta_I \cos^2\theta_R B_{00}[m_Z^2, m_{\zeta_{1R}}^2,
    m_{\zeta_{2I}}^2] \nonumber\\ && + 
 \sin^2\theta_I \sin^2\theta_R B_{00}[m_Z^2, m_{\zeta_{2R}}^2, m_{\zeta_{2I}}^2]) \; ,
\end{eqnarray}
\begin{eqnarray}
\Delta T&=&\frac{-1}{4\pi m_W^2 s_W^2}
(2 s_W^4 A_0[m_{\eta^\pm}^2] + 
\cos^2\theta_I  \sin^2\theta_R B_{00}[0, m_{\zeta_{1I}}^2, m_{\zeta_{2R}}^2]
 - \nonumber\\ && 
 \cos^2\theta_I B_{00}[0, m_{\eta^\pm}^2, m_{\zeta_{1I}}^2] - 
\sin^2\theta_I B_{00}[0, m_{\eta^\pm}^2, m_{\zeta_{2I}}^2] \nonumber\\ && + 
(1-4 s_W^4) B_{00}[0, m_{\eta^\pm}^2, m_{\eta^\pm}^2] - 
\sin^2\theta_R B_{00}[0, m_{\eta^\pm}^2, m_{\zeta_{2R}}^2]\nonumber\\ && + 
\cos^2\theta_R \cos^2\theta_I B_{00}[0, m_{\zeta_{1R}}^2, m_{\zeta_{1I}}^2] + 
\cos^2\theta_R \sin^2\theta_I  B_{00}[0, m_{\zeta_{1R}}^2,
m_{\zeta_{2I}}^2]
\nonumber\\ && - 
\cos^2\theta_R  B_{00}[0, m_{\zeta_{1R}}^2, m_{\eta^\pm}^2] + 
 \sin^2\theta_I \sin^2\theta_R B_{00}[0, m_{\zeta_{2R}}^2, m_{\zeta_{2I}}^2]) \; .
\end{eqnarray}

\end{document}